\documentclass[journal,twoside,web]{ieeecolor}
\usepackage{generic}
\usepackage{cite}
\usepackage{booktabs}
\usepackage{amsmath,amssymb,amsfonts}
\usepackage{algorithmic}
\usepackage{graphicx}
\usepackage{algorithm,algorithmic}
\usepackage{hyperref}
\usepackage{bm}
\usepackage{multirow}
\usepackage{textcomp}
\usepackage{mathptmx}
\usepackage{orcidlink}
\usepackage{mathptmx}
\usepackage{dsfont}
\usepackage{soul}
\usepackage{mathrsfs}
\usepackage{lineno}
\usepackage{flushend}
\hypersetup{hidelinks=true}

\def\BibTeX{{\rm B\kern-.05em{\sc i\kern-.025em b}\kern-.08em
    T\kern-.1667em\lower.7ex\hbox{E}\kern-.125emX}}
\begin{document}
\title{Prototype-Driven and Multi-Expert Integrated Multi-Modal MR Brain Tumor Image Segmentation}
\author{Yafei Zhang$^{\orcidlink{0000-0003-2347-5642}}$, Zhiyuan Li$^{\orcidlink{0009-0009-2804-376X}}$, Huafeng Li$^{\orcidlink{0000-0003-2462-6174}}$, Dapeng Tao$^{\orcidlink{0000-0003-0783-5273}}$, \IEEEmembership{Member, IEEE}
\thanks{This work was supported in part by the National Natural Science Foundation of China, Grant No. 62161015. (Corresponding authours: Huafeng Li and Dapeng Tao)}
\thanks{Y. Zhang, Z. Li and H. Li are affiliated with the Faculty of Information Engineering and Automation, Kunming University of Science and Technology, Kunming 650500, China.(e-mails: zyfeimail@163.com (Y. Zhang); lizy@stu.kust.edu.cn (Z. Li); lhfchina99@kust.edu.cn (H. Li))}
\thanks{D. Tao is with FIST LAB, School of Information Science and Engineering, Yunnan University, Kunming 650091, China (e-mail:dapeng.tao@gmail.com (D. Tao)).}
}
\maketitle
\begin{abstract}
    For multi-modal magnetic resonance (MR) brain tumor image segmentation, current methods usually directly extract the discriminative features from input images for tumor sub-region category determination and localization. However, the impact of information aliasing caused by the mutual inclusion of tumor sub-regions is often ignored. Moreover, existing methods usually do not take tailored efforts to highlight the single tumor sub-region features. To this end, a multi-modal MR brain tumor segmentation method with tumor prototype-driven and multi-expert integration is proposed. It could highlight the features of each tumor sub-region under the guidance of tumor prototypes. Specifically, to obtain the prototypes with complete information, we propose a mutual transmission mechanism to transfer different modal features to each other to address the issues raised by insufficient information on single-modal features. Furthermore, we devise a prototype-driven feature representation and fusion method with the learned prototypes, which implants the prototypes into tumor features and generates corresponding activation maps. With the activation maps, the sub-region features consistent with the prototype category can be highlighted. A key information enhancement and fusion strategy with multi-expert integration is designed to further improve the segmentation performance. The strategy can integrate the features from different layers of the extra feature extraction network and the features highlighted by the prototypes. Experimental results on three competition brain tumor segmentation datasets prove the superiority of the proposed method. The source code will be available at \underline{https://github.com/Linzy0227/PDMINet}.
\end{abstract}


\section{Introduction}
\label{sec:introduction}
\begin{figure}[!t]
	\centering
	\includegraphics[width=0.95\linewidth]{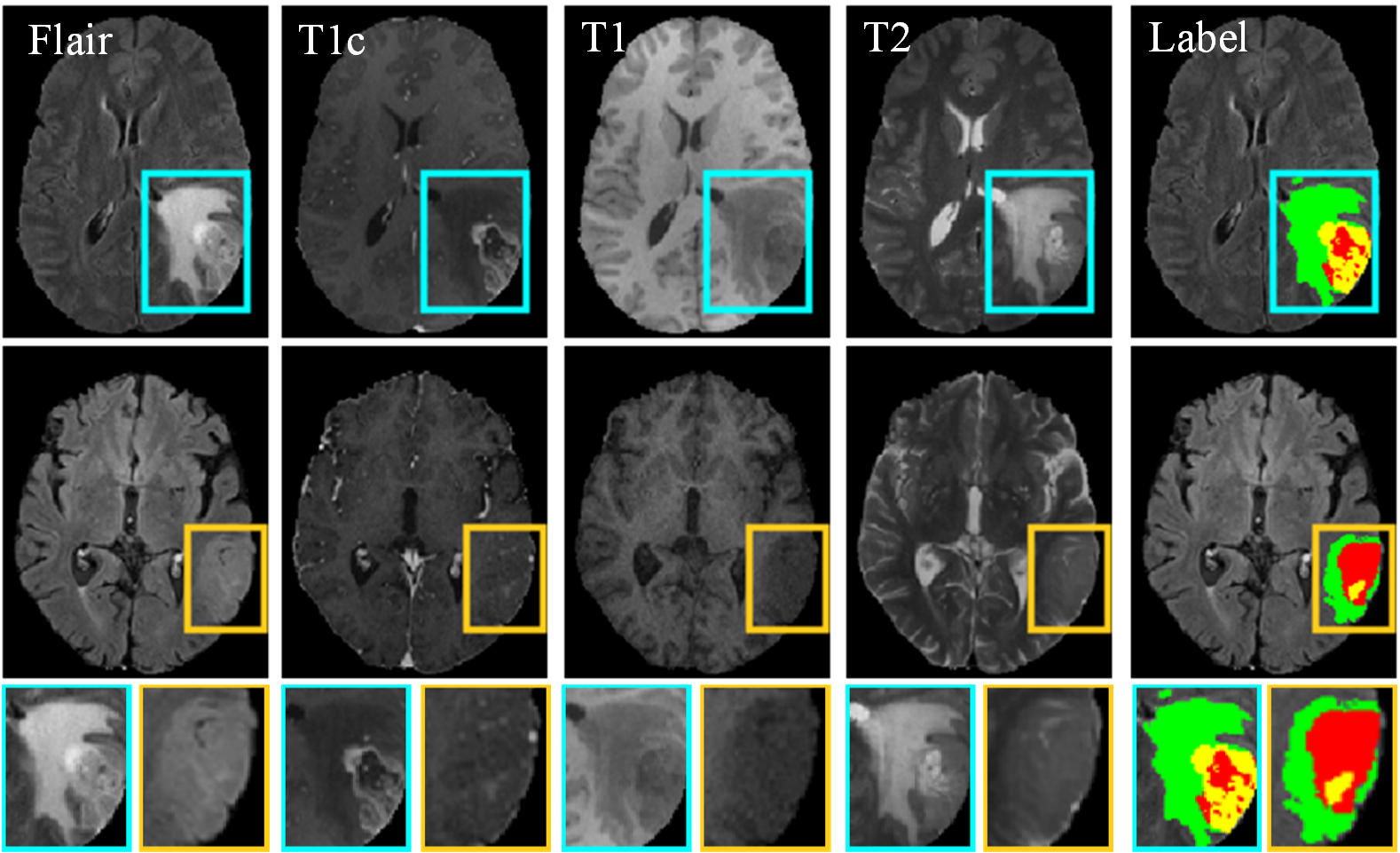}
	\caption{Sensitivity Analysis. The figure exhibits four modalities across two distinct cases. Insight from the locally enlarged detail views reveals how these varied modalities distinctly spotlight different tumor zones. The red, green, and yellow areas respectively denote the necrotic and non-enhancing tumor core (NCR/NET), the peritumoral edema (ED), and  the GD-enhancing tumor (ET).}
	\label{Fig1}
\end{figure}

\IEEEPARstart{T}{he} widespread medical imaging technique known as multi-modal magnetic resonance (MR) imaging is frequently employed in clinical settings. It offers high-resolution visualization of soft tissue lesions, serving as a valuable source of supplementary information for examining brain tumors. Frequently utilized MR sequences for these tumors encompass four modalities: Fluid Attenuated Inversion Recovery (FLAIR), contrast-enhanced T1-weighted (T1c), T1-weighted (T1), and T2-weighted (T2). As depicted in Fig.~\ref{Fig1}, MR images of varying modalities exhibit diverse sensitivities to specific tumor areas. This suggests that different modality brain tumor images possess a degree of complementarity. Such an attribute can be employed to decrease the single-modality uncertainty, thereby enhancing the precision of segmentation essential for clinical diagnosis.

In light of the above reasons, a large number of multi-modal MR brain tumor image segmentation methods have emerged~\cite{Ding2017RFNet, Bauer2011Fully, 9585643, Chen2019S3D, 9793692, Zhuang2023A}. These mainly include traditional machine learning (TML)- and deep learning (DL)-based methods. In TML, segmentation algorithms commonly use support vector machines~\cite{Bauer2011Fully}, random forests~\cite{Zikic2012Decision}, and graph theory~\cite{Saueressig2020Exploring}. Such methods show weak segmentation performance as they need improvement in mining potential statistical characteristics from large samples. In contrast, DL-based methods can effectively solve this problem. Among DL-based methods, the CNN-based method is the most popular, which includes widely used S3D-UNet~\cite{Chen2019S3D}, 3D-SW-Net~\cite{Sun2020A}, ACMINet~\cite{Zhuang2023A}, SegResNet~\cite{Myronenko20193D}, and HPU-Net~\cite{Kong2018Hybrid}. However, since CNNs cannot capture long-distance feature relationships, Transformer-based medical image segmentation methods have been proposed~\cite{Wang2021TransBTS, Hatamizadeh2022Swin, Hatamizadeh2022UNETR}. Although the aforementioned methods can synthesize the complementary information of multi-modal brain tumor images to perform tumor identification and localization, they do not consider the relationship between different modal images. Zhou et al.\cite{Zhou2022A} proposed a triple attention fusion guided segmentation network to address these problems. However, it failed to consider that different tumor regions are mutually encompassing\cite{Zhang2020Exploring} (as shown in Fig. \ref{Fig1}). This leads to information aliasing between different tumor sub-regions, making it extremely difficult to achieve targeted enhancement of different tumor sub-region features.

Moreover, as shown in Fig.~\ref{Fig1}, features of the same category of tumor sub-regions exhibit diversity in the same modality, presenting another challenge for feature-based tumor segmentation. To address these problems, Zhang et al.~\cite{Zhang2020Exploring} proposed a task-structured brain tumor segmentation method rooted in the basic rules of clinical practice. Nevertheless, this method does not sufficiently focus on identifying and locating different tumor sub-regions. Although the fused multi-modal features can offset the limitations of single-modal features, they still fall short of emphasizing the characteristics of tumor sub-regions in a targeted manner, leading to imprecise localizations. This study proposes a prototype-driven and multi-expert integrated tumor segmentation method,  aiming to accentuate tumor-specific characteristics and suppress irrelevant information under the guidance of constructed tumor prototypes. Consequently, the categories of brain tumor sub-regions can be accurately identified and located. In addition, features from different layers of the network are utilized to further aid in the identification and localization of tumor sub-regions.

The proposed method is primarily composed of three parts: Construction of Tumor Prototype (CTP), Prototype-driven Feature Representation and Fusion (PFRF), and Key Information Integration of Multi-expert Integration (KIIMI). In CTP, a multi-modal information interaction mechanism is devised, considering the limitations of information carried by single-modal features. This mechanism realizes the mutual transmission between different modal features and constructs a prototype for each tumor sub-region based on the interacted features. To highlight features related to a specific tumor sub-region, we use such a prototype to enhance feature discrimination. In particular, we incorporate the prototype features into tumor features. Following this, we generate corresponding activation maps to underline features consistent with the prototype category. This approach enables the proposed method to highlight the role of features from individual tumor sub-regions effectively. Moreover, it can suppress the interference of irrelevant features, significantly alleviating the challenges caused by feature aliasing in different regions. Additionally, to further improve the segmentation performance, we embed expert networks with different parameters into various depth layers of the extra feature extraction network. This addition directly assists the identification and localization of tumor regions.

The main contributions are summarized as follows.
\begin{itemize}
	\item A tumor prototype-driven and multi-expert integration method is proposed for brain tumor segmentation. This method not only highlights the significance of features from each tumor sub-region during segmentation but also mitigates the interference caused by the overlapping of sub-regions, thereby enhancing feature discrimination.
	
	\item An information interaction mechanism is developed and the tumor prototype is constructed. With the constructed prototype, we devise a prototype-driven feature representation and fusion method in which the feature discrimination of each tumor sub-region is enhanced under the prototype's guidance.
	
	\item A KIIMI method is developed with the help of an extra feature extraction network. This method integrates the output of the features by different layers and the features highlighted by the prototypes. The integrated results are used for the identification and localization of tumor regions.
\end{itemize}

The rest of this paper is organized as follows. Section~\ref{sec:related} discusses the related state-of-the-art works. Section~\ref{sec:method} elaborates the proposed method in detail. The experimental results are explained in Section~\ref{sec:experiment} and Section~\ref{sec:conclusion} summarizes the content of this paper and draws conclusions.

\section{Related Work}
\label{sec:related}

\subsection{Brain tumor segmentation}
Image segmentation is used to transform the representation of an image into something more meaningful and easier to analyze. Medical image segmentation, in particular, is beneficial for auxiliary diagnosis. However, MR images often suffer from low contrast and a low signal-to-noise ratio due to the limitations of image acquisition equipment, complicating the task of brain tumor segmentation. At present, brain tumor segmentation methods can be roughly categorized into three types~\cite{Havaei2017Brain, Chen2018Focus, Wang2019Automatic, Rezaei2019voxel, Hatamizadeh2022UNETR, Xing2022NestedFormer}: methods based on convolutional neural networks (CNNs), methods based on generative models, and methods that combine Transformers with CNNs.

In the realm of brain tumor segmentation, CNNs represent the most prevalent deep learning methodology. Havaei et al.~\cite{Havaei2017Brain}, in particular, proposed a CNN-based dual-branch model to extract both local and global contextual information from brain images. Nevertheless, the network's input being a 2D slice limits its ability to exploit spatial semantic information from volumetric data. Kamnitsas et al.~\cite{Kamnitsas2017Efficient} utilized a multi-scale 3D CNN to extract lesion features, complemented by a fully connected conditional random field (CRF) to optimize segmentation results. However, the CRF requires reconfiguration when adapting the network to new tasks. In response to the class imbalance in brain tumors, Chen et al.~\cite{Chen2018Focus} proposed cascaded networks that obtain hierarchical information between tumor regions, transforming multi-classification problems into binary classifications to facilitate a coarse-to-fine segmentation. This strategy, nonetheless, can only segment specific tumor regions in a single run, thereby overlooking the correlations amongst different regions.

\begin{figure*}[!t]
	\centering
	\centerline{\includegraphics[height=3.5in,width=6.5in,angle=0]{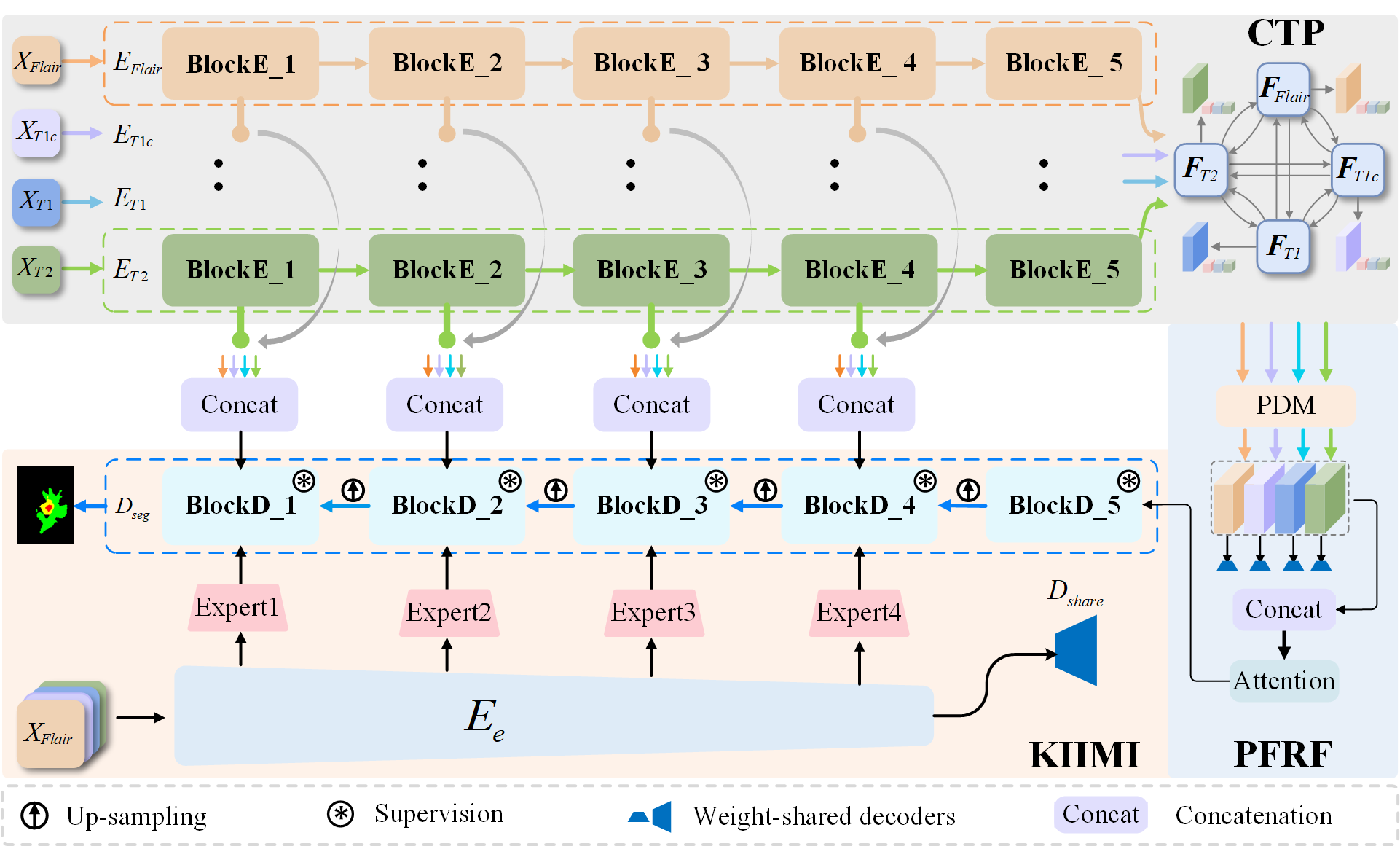}}
	\caption{Proposed prototype-driven multi-modal brain tumor segmentation framework. The encoder, denoted as $E_{\text{m}}$ ($m\in{\text{Flair}, \text{T1c}, \text{T1}, \text{T2}}$), is responsible for extracting features from the $m$-th image modality. Within the encoder $E_{\text{m}}$, the $i$-th feature extraction block is denoted as BlockE$\_{i}$. The decoder part, represented as $D_{\text{seg}}$, also includes several feature extraction blocks labeled as Block$D\_{i}$, where $i$ is the block index. Both encoder and decoder align with the U-Net architecture, each containing $5$ blocks. $D_{\text{share}}$, another decoder, possesses the same structure as $D_{\text{seg}}$, but operates with a distinct set of parameters.}
	\label{Fig2}
\end{figure*}

Generative model-based methods mainly include segmentation methods based on generative adversarial networks (GAN) and auto-encoders~\cite{Rezaei2019voxel, Myronenko20193D}. Methods based on GAN use a segmentation network to produce a segmentation result, and the discriminator distinguishes the true or false of the segmented results~\cite{Rezaei2019voxel}. Although such methods can improve segmentation performance to a certain extent, GAN limits the improvement space due to the convergence difficulty. Auto-encoder-based methods usually add an auto-encoder branch to the decoder branch of U-Net to regularize U-Net encoder~\cite{Myronenko20193D}, thereby improving the feature extraction ability of the network. However, the different branch makes it more challenging to train the network.

Since CNNs cannot mine the long-distance relationship of features, brain tumor segmentation based on the combination of CNNs and Transformers~\cite{Vaswani2017Attention} has become popular~\cite{ Hatamizadeh2022UNETR, Hatamizadeh2022Swin, Xing2022NestedFormer}. Particularly, Hatamizadeh et al.~\cite{Hatamizadeh2022UNETR, Hatamizadeh2022Swin} used Transformer as an encoder to capture global multi-scale features while employing the CNN-based decoder to segment brain tumors. It is not designed for the extensive use of multi-modal information and also needs to pay attention to the global correlation within and between modalities. To address this problem, Xing et al.~\cite{Xing2022NestedFormer} introduced a Transformer encoder embedded with modal awareness and a CNN decoder to segment tumor images. However, those methods treat different areas of the tumor uniformly and do not highlight the role of different sub-region features of the tumor, resulting in unsatisfactory segmentation results. To solve the problem, this paper proposes a tumor prototype-driven approach, which can give prominence to the features of each sub-region of tumor in segmentation, and promotes the identification and localization of these regions.

\subsection{Feature fusion and interaction in brain tumor segmentation}
Image fusion is the process of combining multiple images of the same scene, taken either by different sensors or the same sensor under varying parameter settings, into a single image. Fused images can effectively enhance the accuracy and comprehensiveness of object descriptions. Various fusion strategies have been proposed to exploit the complementary information in multi-modal brain tumor data~\cite{Wang2021TransBTS, Chen2019OctopusNet, Zhang2021Cross, Zhou2021Latent}. For instance, Wang et al.\cite{Wang2021TransBTS} directly integrated images from different modalities and inputted them into a single network for feature extraction. However, they did not consider the importance of different modal features, leading to insufficient mining of complementary information. Chen et al.\cite{Chen2019OctopusNet} and Zhou et al.\cite{Zhou2021Latent} proposed a hierarchical fusion strategy that integrates complementary information by fusing features extracted by a specific encoder at different stages, and then forwards the fused result to a decoder for segmentation via a skip connection. Nevertheless, these fusion strategies overlook the dependencies between modes, thus limiting the segmentation performance. In contrast, Xing et al.\cite{Xing2022NestedFormer} utilized self-attention and cross-attention to calculate the global relationship between modalities, enabling feature extraction from relationship-embedded complementary modalities. Additionally, Zhang et al. \cite{Zhang2022mmFormer} introduced a specific modal encoder and a modal-related encoder to establish the relationship between modalities and to obtain modal invariant features by explicitly aligning the correlation between different modalities. Lastly, Zhou et al.\cite{Zhou2022A} adopted a tri-attention fusion strategy in which the modal-specific features extracted by specific encoders are weighted and fused based on channel, spatial, and channel-spatial information to highlight the most relevant features of the tumor region in different modalities.

Since different modalities often focus on varying tumor regions, many methods tend to overlook the specific contributions of each modality. Zhang et al.\cite{Zhang2021Modality} adopted a modal-aware fusion strategy, which generates a set of weights for each modality in a learnable manner, allowing for the adaptive aggregation of different modal-specific features. Meanwhile, Zhang et al.\cite{Zhang2021Cross} introduced a cross-modal deep feature learning framework based on GAN. This approach facilitates the comprehensive utilization of multi-modal features through cross-modal feature transformation and fusion. Contrary to the aforementioned methods, this paper designs a feature interaction mechanism to construct tumor prototypes. This mechanism enables the mutual exchange of information across different modalities via cross-representation, setting the groundwork for constructing specific tumor region prototypes enriched with comprehensive information.

\section{Proposed Method}
\label{sec:method}

\subsection{Overview}
The proposed prototype-driven multi-modal brain tumor segmentation framework is shown in Fig.~\ref{Fig2}. CTP is used to construct feature prototypes describing different types of tumor sub-regions to guide the localization of the sub-regions. With the aid of tumor prototypes, PFRF is used to highlight the features of each sub-region of the tumor. KIIMI integrates complementary features on different depth layers and improves feature representation ability.

\subsection{Construction of Tumor Prototype}
\begin{figure}[!t]
	\centering
	\includegraphics[width=0.95\linewidth]{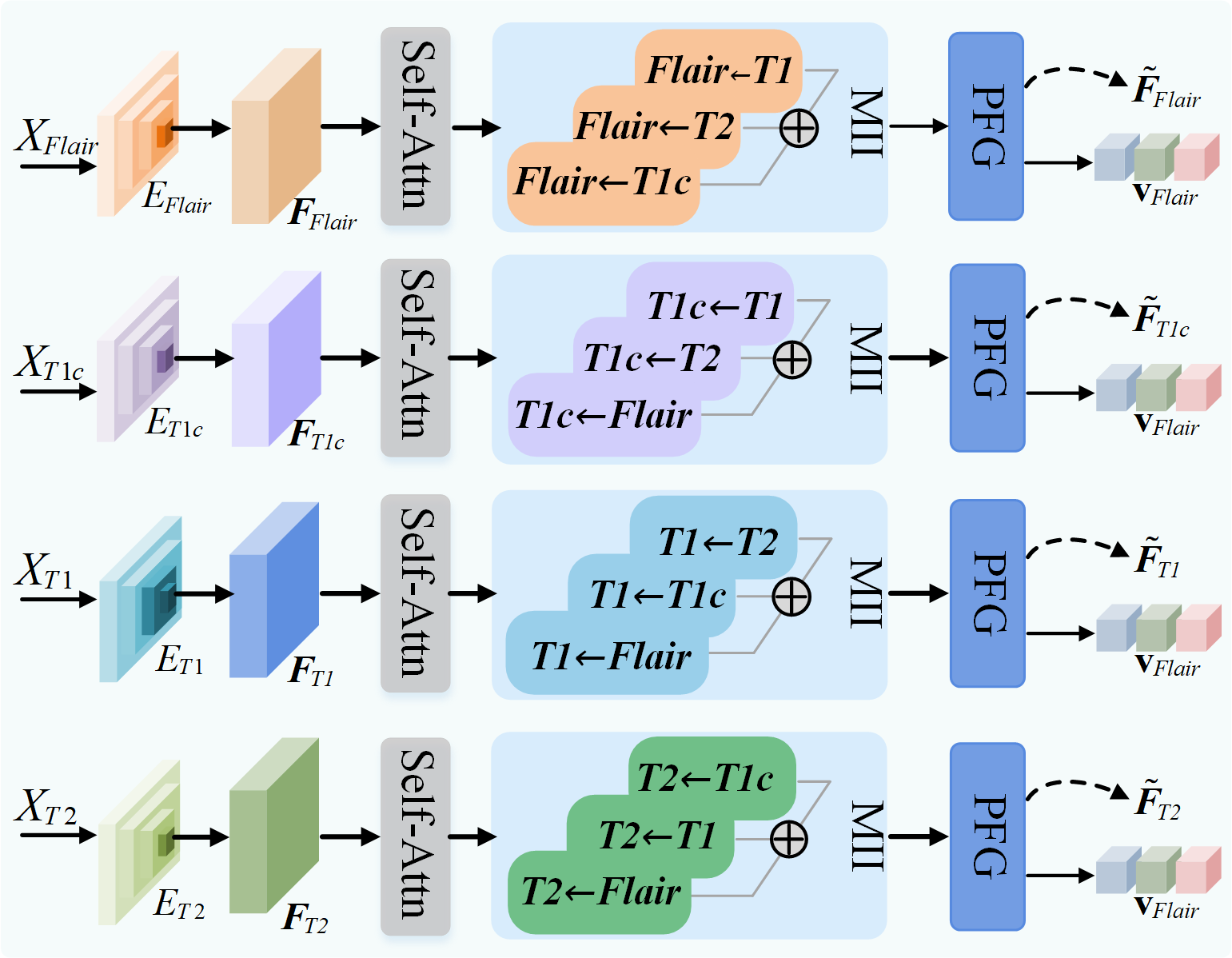}
	\caption{Construction of tumor prototype (CTP).}
	\label{Fig3}
\end{figure}

Different modalities show different sensitivities to different regions of brain tumors and present different features, while similar tumor regions show diversity in multi-modal images, which brings challenges to the localization and classification of tumor sub-regions. To solve this problem, we use tumor prototypes to drive tumor localization. To obtain an information-complete tumor prototype, CTP is devised, as shown in Fig.~\ref{Fig3}, which consists of Multi-modal Information Interaction (MII) and Prototype Feature Generation (PFG).

\begin{figure}[!t]
	\centering
	\includegraphics[width=0.95\linewidth]{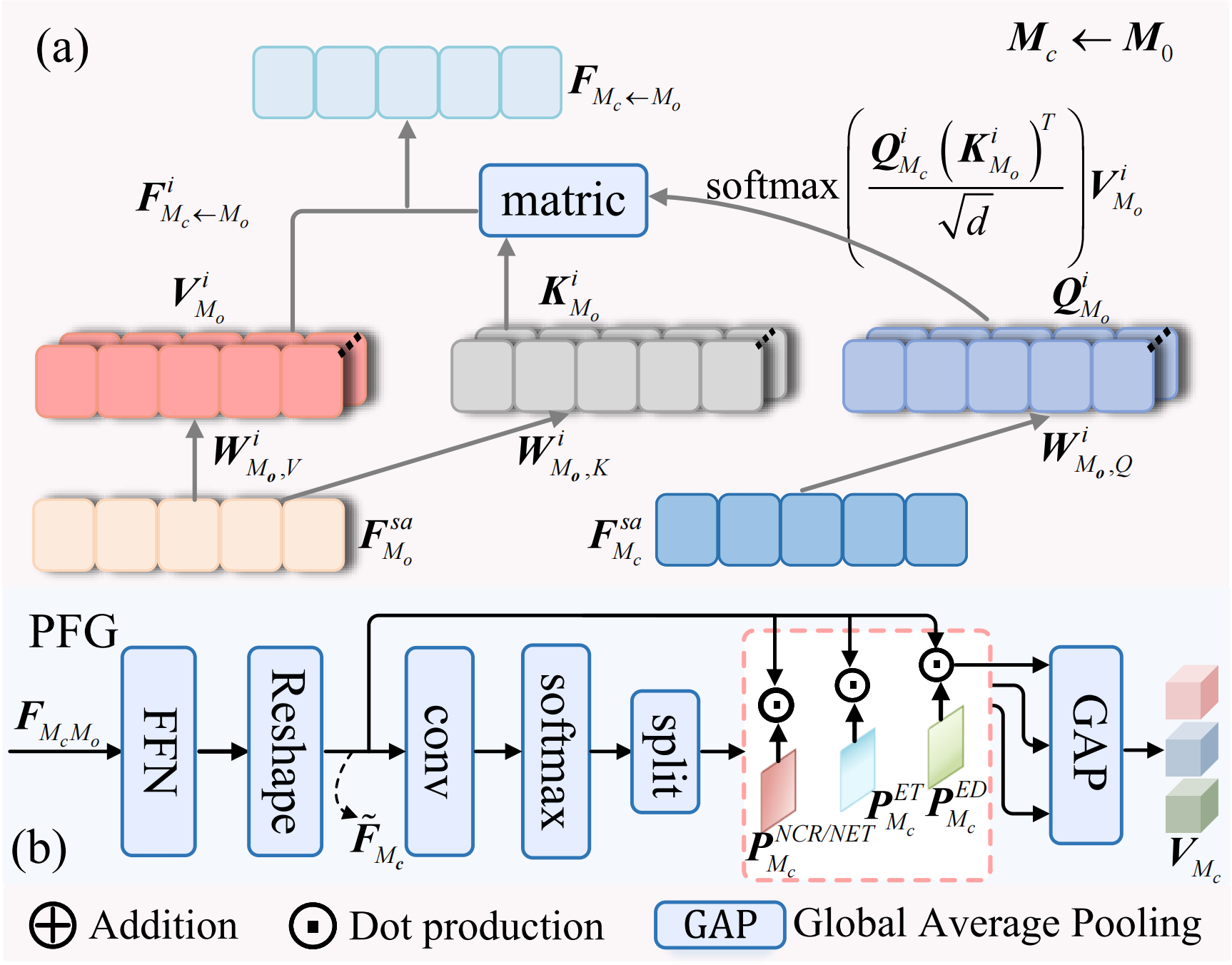}
	\caption{Information interaction and prototype feature generation (PFG)
		framework.}
	\label{Fig4}
\end{figure}

In order to extract modal-unique features, we use encoders with not shared parameters for different modalities, and the U-Net structure is adopted to design encoders $E_{m}$ and decoder $D_{seg}$. Each encoder $E_{m}$ contains $5$ feature extraction blocks, and each block consists of 3D convolution, instance normalization, and LeakyReLU. Given a brain tumor image	$\bm X_{m} \in \mathbb{R}^{1\times H\times\ W\times D}$, the features extracted by encoder $E_{m}$ are $\bm F_{m}\in \mathbb{R}^{C\times H\times\ W\times D}$,
where $m\in\{Flair, T1c, T1, T2\}$ represents the modality of $\bm X_{m}$. $H$ and $W$ are the height and width of the input image, and $D$ is the number of slices. Since the features of specific regions in a tumor are not complete in a single modality, it is difficult to obtain a category prototype of a tumor region with complete information from a uni-modality tumor image. Therefore, an MII mechanism is designed in CTP, as shown in Fig.~\ref{Fig4}(a).

The MII realizes mutual feature transmission through an interactive representation of different modal features, compensating for the impact of incomplete category information of single-modal features on the prototype construction. Given the current modality $M_{c}\in\{Flair, T1c, T1, T2\}$, $M_{o}\in\{Flair, T1c, T1, T2\}$ is other modality except for $M_{c}$, and $M_{c} \bigcap M_{o}= \emptyset$. $M_{c}\leftarrow M_{o}$ denotes information transfer from $M_{o}$ to $M_{c}$. We employ cross-attention to facilitate information exchange across different modalities. Let the feature obtained after $\bm F_{M_{o}}$ passing through the self-attention layer be $\bm F_{M_{o}}^{sa}\in \mathbb{R}^{HWD\times C'}$. In MII, the process of $M_{c}\leftarrow M_{o}$ can be expressed as:
\begin{equation}
	\bm F_{M_{c}\leftarrow M_{o}}^{i} = \textrm{softmax}\left(\frac{\bm Q_{M_{c}}^i (\bm K_{M_{o}}^i)^T}{\sqrt{d}} \right) \bm V_{M_{o}}^i
\end{equation}
where $\bm F_{M_{c}\leftarrow M_{o}}^{i}$ represents the outcome of the $i$-th attention head in $M_{c}\leftarrow M_{o}$. $\bm Q_{M_{c}}^i=LN(\bm F_{M_{c}}^{sa})\bm W_{M_{c},Q}^i$, $\bm K_{M_{o}}^i=LN(\bm F_{M_{o}}^{sa})\bm W_{M_{o},K}^i$, $\bm V_{M_{o}}^i=LN(\bm V_{M_{o}}^{sa})\bm W_{M_{o},V}^i$. $\bm W_{k,l}^i\in\mathbb{R}^{C'\times d} (k=M_{c},M_{o}, l={\bm Q,\bm K,\bm V})$ denotes the parameter matrix of the linear mapping and $LN$ stands for layer normalization. Following the acquisition of features from the $i$-th attention head, denoted as $\bm F_{M_{c}\leftarrow M_{o}}^{i}$, the comprehensive formulation for $\bm F_{M_{c}\leftarrow M_{o}}$ is presented as:
\begin{equation}
	\bm F_{M_{c}\leftarrow M_{o}} = [\bm F_{M_{c}\leftarrow M_{o}}^{1},\bm F_{M_{c}\leftarrow M_{o}}^{2},\cdots,\bm F_{M_{c}\leftarrow M_{o}}^{N_{h}}]\bm W_{M_{c}\leftarrow M_{o}}
\end{equation}
where $\left[\cdot,\cdot\right] $ denotes concatenation operation. $\bm W_{M_{c}\leftarrow M_{o}}$is the linear mapping matrix. $N_{h}$ is the number of heads, which is set to $8$ according to \cite{Zhang2022mmFormer}. Once $M_{o}$ has traversed all the modalities except $M_{c}$, it can realize the interaction of other modal information with $M_{c}$. The integrated features of $M_{o}$ and $M_{c}$, which contain the information of all modalities, can be expressed as:
\begin{equation}
	\bm F_{M_{c}M_{o}} = \sum_{i=1}^{3}\bm F_{M_{c}\leftarrow M_{o}^i} + \bm F_{M_{c}}
\end{equation}
where $M_{o}^1$, $M_{o}^2$ and $M_{o}^3$ are three modalities except $M_{c}$ that differ from each other.

After $\bm F_{M_{c}}$ interacts with other modal features, $\bm F_{M_{c}M_{o}}$ is injected with information from other modalities. In PFG shown in Fig.~\ref{Fig4}(b), we can effectively avoid incomplete prediction of tumor prototype with $\bm F_{M_{c}M_{o}}$. Technically, $\bm F_{M_{c}M_{o}}$ goes through a Feed Forward Network (FFN) consisting of linear layers, activation function GELU, and Dropout for feature extraction, and then through Reshape to obtain $\tilde{\bm F}_{M_{c}}\in\mathbb{R}^{C'\times H\times W\times D}$. $\bm P_{M_{c}}^{NCR/NET}$, $\bm P_{M_{c}}^{ED}$ and $\bm P_{M_{c}}^{ET}$ are probability maps representing different region categories of tumor obtained by $\tilde{\bm F}_{M_{c}}$ after convolution, softmax and split:
\begin{equation}
	\{\bm P_{M_{c}}^{NCR/NET},\bm P_{M_{c}}^{ED},\bm P_{M_{c}}^{ET}\} = \text{split}(\text{softmax}(\text{conv}(\tilde{\bm F}_{M_{c}})))
\end{equation}
where $NCR/NET$ is the necrotic and non-enhancing tumor core ($NCR/NET$), $ED$ stands for the peritumoral edema, and $ET$ is the GD-enhancing tumor.
The obtained prototypes in different regions of current modal tumors can be described by:
\begin{equation}
	\bm v_{M_{c}}^i = \frac{\sum_{d=1}^{D}\sum_{w=1}^{W}\sum_{k=1}^{H}(\tilde{\bm F}_{M_{c}}\odot \bm P_{M_{c}}^i)_{k,w,d}}{H\times W\times D}
\end{equation}
where $i\in\{NCR/NET, ED, ET\}$, $\odot$ is the dot production.
\subsection{Prototype-driven Feature Representation and Fusion}
\begin{figure}[!t]
	\centering
	\includegraphics[width=0.95\linewidth]{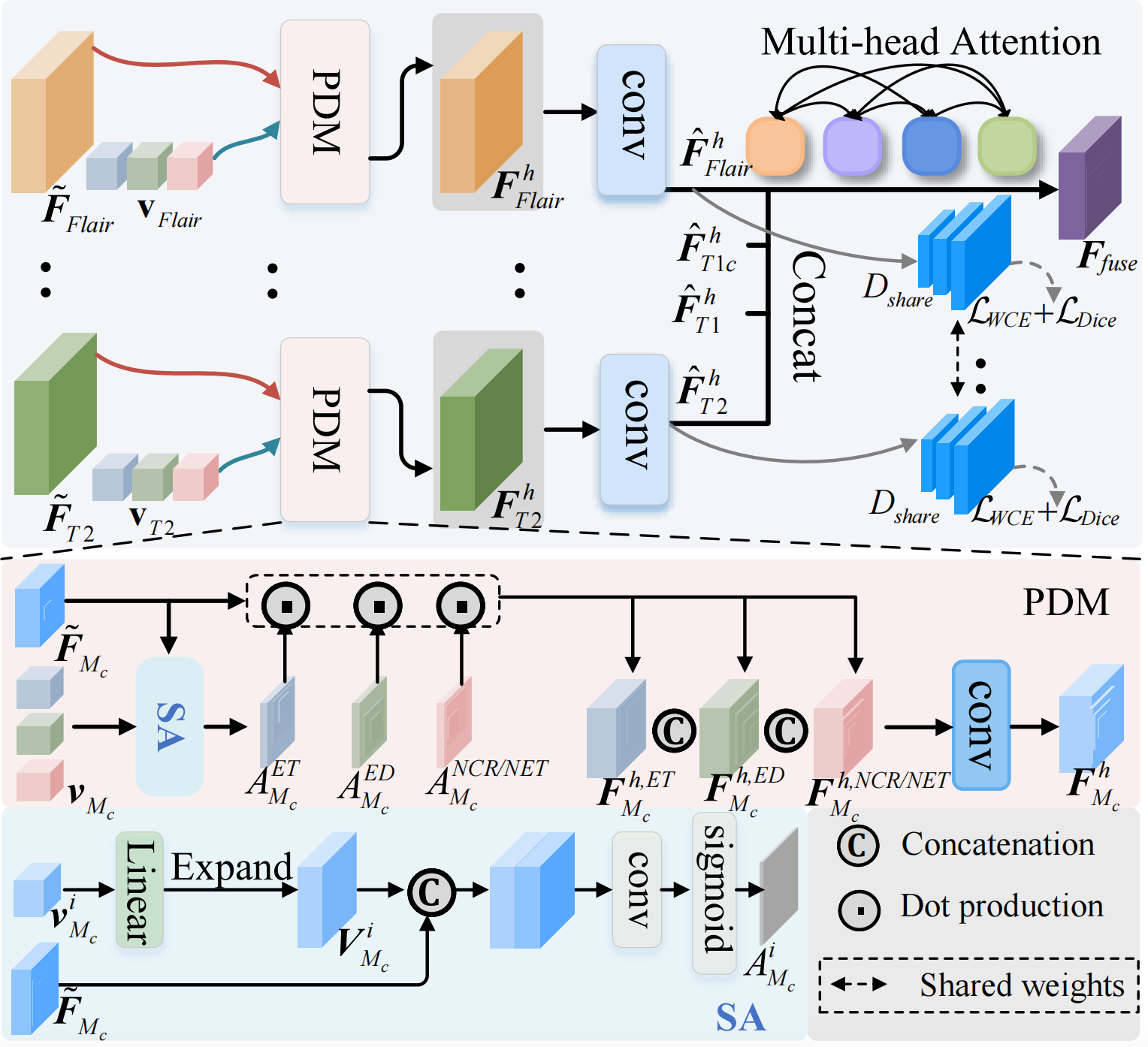}
	\caption{Prototype-driven feature representation and fusion (PFRF) framework. PDM denotes the prototype drive module..}
	\label{Fig5}
\end{figure}

In order to highlight the features of tumor sub-regions in each modality, a PFRF module included PDM  is designed, as shown in Fig.~\ref{Fig5}. In PDM, we first obtain a matrix representation $\bm V_{M_{c}}^i$ of the same dimension with $\tilde{\bm F}_{M_{c}}$ through linear mapping and expansion operations on $\bm v_{M_{c}}^i$, and concatenate $\bm V_{M_{c}}^i$ and $\tilde{\bm F}_{M_{c}}$ on channels to strengthen their common tumor features. After that, we apply a $1 \times 1\times 1$ convolutional layer and ReLU on the concatenated feature to generate the feature activation map $\bm A_{M_{c}}^i$. The value in $\bm A_{M_{c}}^i$ reflects the extent to which the features at the same location of $\tilde{\bm F}_{M_{c}}$ belonging to $i$-th ($i\in\{NCR/NET, ED, ET\}$) category of the tumor sub-regions. Thus, according to (6), the features of the $i$-th category region of the tumor can be highlighted:
\begin{equation}
	\bm F_{M_{c}}^{h,i} = \tilde{\bm F}_{M_{c}}\odot \bm A_{M_{c}}^i
\end{equation}

After $\bm F_{M_{c}}^{h, NCR/NET}$, $\bm F_{M_{c}}^{h, ED}$, $\bm F_{M_{c}}^{h, ET}$ are generated, we apply concatenation and convolution on them to obtain complete tumor feature representation of current modal:
\begin{equation}
	\bm F_{M_{c}}^{h} = \textrm{conv}([\bm F_{M_{c}}^{h, NCR/NET}, \bm F_{M_{c}}^{h, ED}, \bm F_{M_{c}}^{h, ET}])
\end{equation}
Finally, $1\times1\times1$ convolutional layer on $\bm F_{M_{c}}^{h}$ is applied to integrate the features and $ \hat{\bm F}_{M_{c}}^{h}$ is obtained. Features of different modalities are concatenated together and passed through a multi-head attention layer to obtain the fused features:
\begin{equation}
	\bm F_{fuse} = \text{MHSA}([\hat{\bm F}_{Flair}^{h}, \hat{\bm F}_{T1c}^{h}, \hat{\bm F}_{T1}^{h}, \hat{\bm F}_{T2}^{h}])
\end{equation}
where MHSA is the multi-head self-attention.

\subsection{Multi-expert Key Information Integration}
The features extracted from different layers have different characteristics. The semantic information carried by deeper features is richer, while the shallow features have ample underlying visual information. A KIIMI method is proposed to effectively use the information to identify and localize tumor sub-regions. In view of the complex training for a specific modal encoder, it will not be conducive to improving encoder capabilities if the above tasks are directly realized on $E_{m}$. To this end, we introduce an extra encoder $E_{e}$ to highlight the role of features at different layers in locating and classifying brain tumor regions. The structure of $E_{e}$ is the same as $E_{m}$, all made up of 5 encoder blocks. We put the concatenated result of brain tumor images from all modalities into $E_{e}$, and the output of the $l$-th block is $\bm F_{e}^l$. Besides, we put $\bm F_{e}^l$ to the $l$-th expert network to learn features of $3$ tumor regions, i.e., $NCR/NET$, $ED$ and $ET$. In the decoding stage, we concatenate the tumor features learned by the expert network with the features in encoder $E_{m}$ to assist the segmentation of brain tumors, as shown in Fig.~\ref{Fig2}. It can be seen that we do not perform the above operations on the output features of the last layer network because the features that PFRF inputs to BlockD\_$5$ already have rich semantic information.
\begin{figure}[!t]
	\centering
	\includegraphics[width=\linewidth]{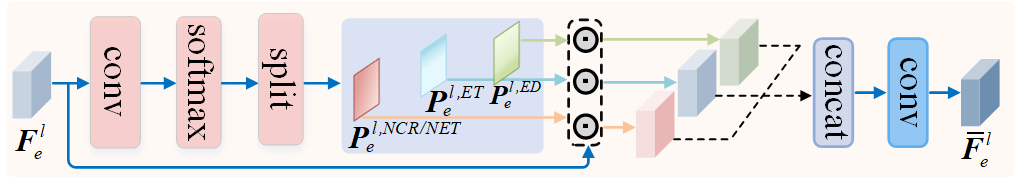}
	\caption{Expert network structure.}
	\label{Fig6}
\end{figure}

In KIIMI, we use experts with the same structure but different parameters for feature extraction, which is shown in Fig.~\ref{Fig6}. Each expert consists of convolutional layers, softmax, and split operations. To extract features of three tumor sub-regions, probability maps $\bm P_{e}^{l,NCR/NET}$, $\bm P_{e}^{l,ED}$, $\bm P_{e}^{l,ET}$ representing categories of different sub-regions in tumor are obtained after convolution, softmax and split on $\bm F_{e}^l$:
\begin{equation}
	\{\bm P_{e}^{l,NCR/NET},\bm P_{e}^{l,ED},\bm P_{e}^{l,ET}\}=\textrm{split}(\textrm{softmax}(\textrm{conv}(\bm F_{e}^l)))
\end{equation}

According to $\bm P_{e}^{l,NCR/NET}$, $\bm P_{e}^{l,ED}$ and $\bm P_{e}^{l,ET}$, the non-tumor area is restrained and the tumor information is enhanced and integrated by (10):
\begin{equation}
	\bar{\bm F}_{e}^l = \textrm{conv}([\bm F_{e}^l\odot\bm P_{e}^{l,NCR/NET}, \bm F_{e}^l\odot\bm P_{e}^{l,ED},\bm F_{e}^l\odot\bm P_{e}^{l,ET}])
\end{equation}
In decoding stage, we use skip-connection to concatenate $\bar{\bm F}_{e}^l$ and output features of corresponding block in encoder $\{E_{Flair}, E_{T1c}, E_{T1}, E_{T2}\}$ and send them to decoder $D_{seg}$ for final segmentation result.

\subsection{Loss Function}
To ensure the multi-modal segmentation performance, the following loss functions are used to optimize model parameters.

\subsubsection{Loss for CTP}. To make the predicted probability map $\bm P_{M_{c}}^{NCR/NET},\bm P_{M_{c}}^{ED},\bm P_{M_{c}}^{ET}$ show the category of tumor sub-regions, similar to~\cite{Chen2019Robust}, we use weighted cross-entropy loss $\mathcal{L}_{WCE}$ and Dice loss $\mathcal{L}_{Dice}$ to optimize parameters in CTP module. Here, $\mathcal{L}_{Dice}$ is defined as:
\begin{equation}
	\begin{aligned}
		& \mathcal{L}_{Dice}(\uparrow\bm P_{M_{c}}, \bm P)
		=\\
		&1-\sum_{k\in\Omega_{c}}\frac{2\sum_{j=1}^{N}p_{j}^k p_{M_{c},j}^k}{\sum_{j=1}^{N}p_{j}^k p_{j}^k+\sum_{j=1}^{N}p_{M_{c},j}^k p_{M_{c},j}^k+\epsilon}
	\end{aligned}\
\end{equation}
and $\mathcal{L}_{WCE}$ is defined as:
\begin{equation}
	\mathcal{L}_{WCE}(\uparrow\bm P_{M_{c}}, \bm P)=-\sum_{k\in\Omega_{c}}\sum_{j=1}^{N}\omega_{k}\cdot p_{j}^k\log p_{M_{c},j}^k
\end{equation}
where $\Omega_{c}=\{BG(background), NCR/NET, ED, ET\}$. $p_{j}^k$ and $p_{M_{c},j}^k$ represent the ground truth and probability prediction of voxel $j$ on class $k$, respectively. $\omega_{k}$ is the weight of region $k$. $N=H\times W\times D$, $\epsilon= 1\times 10^{-5}$. $\uparrow\bm P_{M_{c}}\in\mathbb{R}^{4\times HWD}$ and $\bm P\in\mathbb{R}^{4\times HWD}$ denote the predicted probability map by PFG in CTP module and ground truth, respectively. $p_{M_{c},j}^k$ and $p_{j}^k$ are the values at the position of row $k$ and column $j$ in $\uparrow{\bm P_{M_{c}}}$ and $\bm P$. $\uparrow$ represents an upsampling operation. In the CTP module, the loss $\mathcal{L}_{ctp}$ is the combined loss of $\mathcal{L}_{WCE}$ and $\mathcal{L}_{Dice}$:
\begin{equation}
	\mathcal{L}_{ctp} =\sum_{M_{c}}\mathcal{L}_{WCE}(\uparrow\bm P_{M_{c}}, \bm P)+\mathcal{L}_{Dice}(\uparrow\bm P_{M_{c}}, \bm P)
\end{equation}

\subsubsection{Loss for $D_{share}$}. In PFRF and KIIMI modules, decoder $D_{share}$ is used for brain tumor segmentation. The loss function $\mathcal{L}_{share}$ is used to constrain the segmentation results of $D_{share}$ such that the encoders $E_{m}$ and $E_{e}$ could sufficiently extract features:
\begin{equation}
	\mathcal{L}_{share} =\sum_{s\in\Omega}\mathcal{L}_{WCE}(\bm P_{s}, \bm P)+\mathcal{L}_{Dice}(\bm P_{s}, \bm P)
\end{equation}
where $\Omega=\{Flair, T1c, T1, T2, Flair\textcopyright T1c\textcopyright T1\textcopyright T2\}$. $\bm P_{Flair}$, $\bm P_{T1c}$, $\bm P_{T1}$ and $\bm P_{T2}$ are the predicted probability maps of $D_{share}$ in PFRF module. $\bm P_{Flair\textcopyright T1c\textcopyright T1\textcopyright T2}$ is the predicted probability map of $D_{share}$ in KIIMI module.

\subsubsection{Loss for $E_{e}$}. The following loss function is used to optimize the parameters of each expert network:
\begin{equation}
	\mathcal{L}_{exp} =\sum_{l=1}^{L}\mathcal{L}_{WCE}(\uparrow\bm P_{e}^{l}, \bm P)+\mathcal{L}_{Dice}(\bm P_{e}^{l}, \bm P)
\end{equation}
where $\bm P_{e}^{l}$ represents the predicted probability map of tumor regions from the output feature of the $l$-th block in encoder $E_{e}$. Here $L$ is taken as $4$, which denotes the first four blocks of encoder $E_{e}$.

\subsubsection{Loss for $D_{seg}$}. During the training phase, we use the following loss on each block of the segmentation decoder $D_{seg}$ to optimize its parameters:
\begin{equation}
	\mathcal{L}_{b} =\mathcal{L}_{WCE}(\uparrow\bm P_{b}, \bm P)+\mathcal{L}_{Dice}(\uparrow\bm P_{b}, \bm P)
\end{equation}
where $b$ is the index of block in $D_{seg}$, and $\bm P_{b}$ is the predicted probability map of the $b$-th Block. The total loss is:
\begin{equation}
	\mathcal{L}_{total} = \mathcal{L}_{ctp}+\mathcal{L}_{share}+\mathcal{L}_{exp}+\sum_{b=1}^{5}\mathcal{L}_{b}
\end{equation}

\section{Experiments Results and Discussions}
\label{sec:experiment}

\subsection{Datasets and evaluation metrics}
To verify the effectiveness of the proposed method, we evaluate performance on three publicly available datasets: BraTS2018, BraTS2019 and BraTS2020~\cite{Menze2015Multimodal}. All cases in each dataset consist of images of four modalities: Flair, T1, T1c, and T2. The tumor region in each modal image contains four categories: $BG$, $NCR/NET$, $ED$ and $ET$. The size of each image is $240\times 240\times 155$. Training sets in BraTS2018 BraTS2019 and BraTS2020 consist of MRI sequences from 285, 335 and 369, cases with known labels, respectively. Validation sets consist of MRI sequences from 66, 125 and 125 cases with unknown labels, respectively. In those datasets, MRI images are skull-stripped, co-registered, and re-sampled to $1mm^{3}$ resolution. In this work, we use the Dice score~\cite{Dice1945Measures} and Hausdorff (HD95) distance to evaluate the segmentation accuracy of the model.
\subsection{Implement details}

Adhering to the procedures specified in~\cite{Dorent2019Hetero}, we perform a normalization process on each MR image within our datasets, ensuring the brain region achieves zero-mean unit variance. We conduct all experiments utilizing two NVIDIA GTX3090 GPUs, under the auspices of the Pytorch 1.12.1 framework~\cite{Paszke2019PyTorch}. During the training phase, we implement data augmentation techniques which include random cropping, random mirror flipping across coronal, sagittal, and axial planes with a probability of $0.5$, applying a random intensity shift within a range of $[-0.1,0.1]$, and adjusting the scale within a range of $[0.9,1.1]$.

During the model training process, we randomly crop the image to a size of $128\times 128\times 128$. Besides, we utilize Adam \cite{Kingma2014Adam} as an optimizer for the model parameters, with $\beta_{1}$ and $\beta_{2}$ set to 0.9 and 0.999, respectively, and the weight decay set at $1 \times 10^{-5}$. For each epoch, the learning rate, denoted by $lr$, is defined as $lr = baseRate \times (1 - \frac{currentEpoch}{totalEpochs})^p$, utilizing the ``poly" learning rate policy for dynamic adjustment, with $p$ being 0.9. The baseline learning rate, $baseRate$, is set at $2 \times 10^{-4}$ and the total number of epochs, $totalEpochs$, is 2000. To further boost model performance during the testing phase, we implement Test Time Augmentation, following the approach outlined in~\cite{Wang2021TransBTS}.

\begin{table*}[htbp]\footnotesize
	\centering {\caption{Comparison of different methods on BraTS2018 validation set. WT represents a region consisting of NCR/NET, ED, and ET. TC represents a region consisting of NCR/NET and ET. ``Avg." denotes the average value in the TC, ET and WT. Bold indicates the best result.}\label{table1} 
		\renewcommand\arraystretch{1.0}
		{\footnotesize\centerline{\tabcolsep=12.0pt
		\begin{tabular*}{0.9\textwidth}{cc cccc cccc}
			\hline
				\multirow{2}*{Methods} & \multicolumn{4}{c}{Dice Score($\%$)$\uparrow$} & \multicolumn{4}{c}{HD95 (mm)$\downarrow$} \\
			\cline{2-5} \cline{7-10}
			&TC &ET &WT  &Avg. &&TC &ET &WT  &Avg.\\
			\hline
			V-Net\cite{Milletari2016VNet,Wang2021TransBTS}    &81.00 &76.60 &89.60  &82.40 &&7.82 &7.21  &6.54  &7.19 \\
			3D U-Net\cite{Ahmed20163D,Wang2021TransBTS} &78.30 &75.40  &87.30 &80.33   &&8.03 &4.56  &5.90  &6.16\\
			Auto-Net\cite{Salehi2017Auto}  &73.30 &70.40  &82.20  &75.30  &&13.91 &9.67  &9.61  &11.06\\
			FSENet\cite{Chen2018Focus} &73.10 &70.70  &84.50 &76.10  &&15.07 &10.39  &11.82  &12.43\\
			BTSPSP\cite{Weninger2019Segmentation} &75.80 &71.20  &88.90  &78.63  &&10.91 &6.28  &6.97  &8.05\\
			S3d-unet.\cite{Chen2019S3D} &83.09 &74.93  &89.35  &82.46  &&7.75 &4.43  &4.72  &5.63\\
			TSBTS\cite{Zhang2020Exploring} &82.40 &78.20  &89.60  &83.40  &&9.27 &3.57  &5.73  &6.19\\
			CANet\cite{Liu2021CANet}  &83.40 &76.70  &89.80  &83.30  &&7.67 &3.86  &6.69  &6.07\\
			TransBTS\cite{Wang2021TransBTS}  &81.15 &79.39  &88.92  &83.15  &&6.92 &3.36  &6.60  &5.63\\
			TA-Net\cite{Zhou2022A} &78.40 &68.80  &87.60  &78.27  &&6.90  &6.55  &10.20 &7.88\\
			AST-Net\cite{Wang2019AST}  &\bf85.00 &79.50  &90.50   &85.00  &&9.20 &2.98  &5.95  &6.04\\
			Ours  &84.60 &\bf{81.06}  &\bf{90.57}  &\bf{85.41} &&\bf 6.88 &\bf{2.85} &\bf{3.97} &\bf{4.57}\\
			\hline
	\end{tabular*}}}}
\end{table*}

\subsection{Comparison with state-of-the-arts}
To verify the effectiveness of the proposed method, we compare it with state-of-the-art segmentation methods on BraTS2018, BraTS2019 and BraTS2020 datasets, respectively. In this process, all the methods are tested on the validation sets of BraTS2018, BraTS2019 and BraTS2020, respectively. We upload the test results to the CBICA online platform (https://ipp.cbica.upenn.edu/) for performance evaluation. Experimental results on BraTS2018  are shown in Table \ref{table1}. It can be seen that Dice score reach $84.60\%$, $81.06\%$ and $90.57\%$ in TC, ET and WT regions in the proposed method, and HD95 distance reach $6.88$ mm, $2.85$ mm and $3.97$ mm, respectively. Compared to AST-Net with suboptimal performance, the proposed method is only slightly deficient in Dice of TC, which proves its effectiveness.

\begin{table*}[htbp]\footnotesize
	\centering {\caption{Comparison of different methods on BraTS2019 validation set. Bold indicates the best result.}\label{table2}
		\renewcommand\arraystretch{1.0}
		{\footnotesize\centerline{\tabcolsep=12.0pt
				\begin{tabular*}{0.9\textwidth}{cc cccc cccc}
					\hline
					\multirow{2}*{Methods} & \multicolumn{4}{c}{Dice Score($\%$)$\uparrow$} & \multicolumn{4}{c}{HD95 (mm)$\downarrow$} \\
					\cline{2-5} \cline{7-10}
					&TC &ET &WT  &Avg. &&TC &ET &WT  &Avg.\\
					\hline
				V-Net\cite{Milletari2016VNet,Chang2023DPAFNet}   &76.56 &73.89  &88.73  &79.73  &&8.71 &6.13  &6.26  &7.03\\
				3D U-Net\cite{Ahmed20163D,Chang2023DPAFNet} &72.48 &70.86 &87.38  &76.91  &&8.72 &5.06  &9.43  &7.74\\
				Attn U-Net\cite{Oktay2018Attention}  &77.20 &75.96  &88.81  &80.66  &&8.26 &5.20  &7.76  &7.07\\
				MC-Net\cite{Li2020Multi} &81.30 &77.10  &88.60  &82.33  &&7.41 &6.03  &6.23  &6.56\\
				BTS-SDP\cite{3du-net19}     &80.70 &73.70  &89.40  &81.27   &&7.36  &5.99 &5.68. &6.34\\
				Swin-Unet\cite{Cao2021Swin,Li2022TransBTSV2} &78.75 &78.49  &89.38  &82.21  &&9.26 &6.93  &7.51  &7.90\\
				TransUNet\cite{Chen2021TransUNet,Li2022TransBTSV2} &78.91 &78.17  &89.48  &82.19  &&7.37 &4.83  &6.67  &6.29\\
				TransBTS\cite{Wang2021TransBTS} &81.94 &\bf{78.93} &90.00 &\bf{83.62}   &&6.05 &3.74  &5.64  &5.14\\	
				KiU-Net\cite{Valanarasu2022KiU} &73.92 &73.21  &87.60  &78.24  &&9.89 &6.32  &8.94  &8.38\\			
				CGA U-Net\cite{Li2022Category} &82.32 &78.83  &89.29  &83.48  &&6.72 &3.26  &5.81  &5.26\\
				Ours  &\bf{82.97} &77.69  &\bf{90.16}  &83.61  &&\bf{5.57} &\bf{2.98} &\bf5.25 &\bf{4.60}\\
				\hline
	\end{tabular*}}}}
\end{table*}

The experimental results of different methods on BraTS2019 dataset are shown in Table \ref{table2}, with the same hyper-parameter settings on BraTS2018 dataset. Nevertheless, the proposed framework still obtain optimal Dice values on TC and WT, with $82.97\%$ and $90.16\%$, respectively. Secondly, HD95 distance in TC, ET and WT regions also reach the best, with $5.57$~mm, $2.98$~mm and $5.25$~mm, respectively, significantly superior to other methods. It is worth noting that in the segmentation task of TC, ET and WT, there is currently no method that can achieve the optimal Dice score and HD95 distance at the same time. However, the proposed method delivers certain effectiveness even without fine-tuning the model parameters on this dataset.

\begin{table*}[htbp]\footnotesize
	\centering {\caption{Comparison of different methods on BraTS2020 validation set. Bold indicates the best result.}\label{table3} 
		\renewcommand\arraystretch{1.0}
		{\footnotesize\centerline{\tabcolsep=12.0pt
				\begin{tabular*}{0.9\textwidth}{cc cccc cccc}
					\hline
					\multirow{2}*{Methods} & \multicolumn{4}{c}{Dice Score($\%$)$\uparrow$} & \multicolumn{4}{c}{HD95 (mm)$\downarrow$} \\
					\cline{2-5} \cline{7-10}
					&TC &ET &WT  &Avg. &&TC &ET &WT  &Avg.\\
					\hline
					Basic V-Net\cite{Milletari2016VNet,Chang2023DPAFNet} &75.26 &61.79  &84.63 &73.89 &&12.18 &47.70  &20.41 &26.76\\
					Deeper V-Net\cite{Milletari2016VNet,Chang2023DPAFNet} &77.90 &68.97  &86.11 &77.66 &&16.15 &43.52  &14.50 &24.72\\				
					3D U-Net\cite{Ahmed20163D,Chang2023DPAFNet} &79.59 &76.07 &89.33 &81.66 &&16.41 &44.91  &23.19 &28.17\\
					3D Residual U-Net\cite{Zhang2018Road}  &76.47 &71.63  &82.46 &76.85  &&13.11 &37.42  &12.34  &20.96\\
					ModGen\cite{Zhou2019ModGen,Liu2023M3AE}  &83.70  &73.20  &89.10  &82.00 &&7.70  &40.00  &7.60  &18.43\\
					HDC-Net\cite{Luo2021HDCNet} &82.50 &77.20  &89.40  &83.03  &&9.13 &27.42  &6.73  &14.43\\
					RFNet\cite{Ding2021RFNet,Liu2023M3AE}  &\bf{83.30} &71.90 &89.80 &81.67 &&9.60  &37.60  &7.00  &18.07\\
					ACN\cite{Wang2021ACN,Liu2023M3AE} &81.10 &70.70  &88.70 &80.17  &&11.30 &43.60  &8.10 &21.00\\
					SMU-Net\cite{Azad2022SMUNet,Liu2023M3AE} &\bf{83.30}  &73.60 &89.80 &82.23  &&7.20  &29.20  &5.80 &14.07\\				
					Ours  &82.15  &\bf{79.04}  &\bf{90.05}  &\bf{83.75} &&\bf{7.08} &\bf{26.57}  &6.80  &\bf{13.48}\\
					\hline
	\end{tabular*}}}}    
\end{table*}

\begin{figure*}[t]
	\centering
	\includegraphics[width=0.9\textwidth]{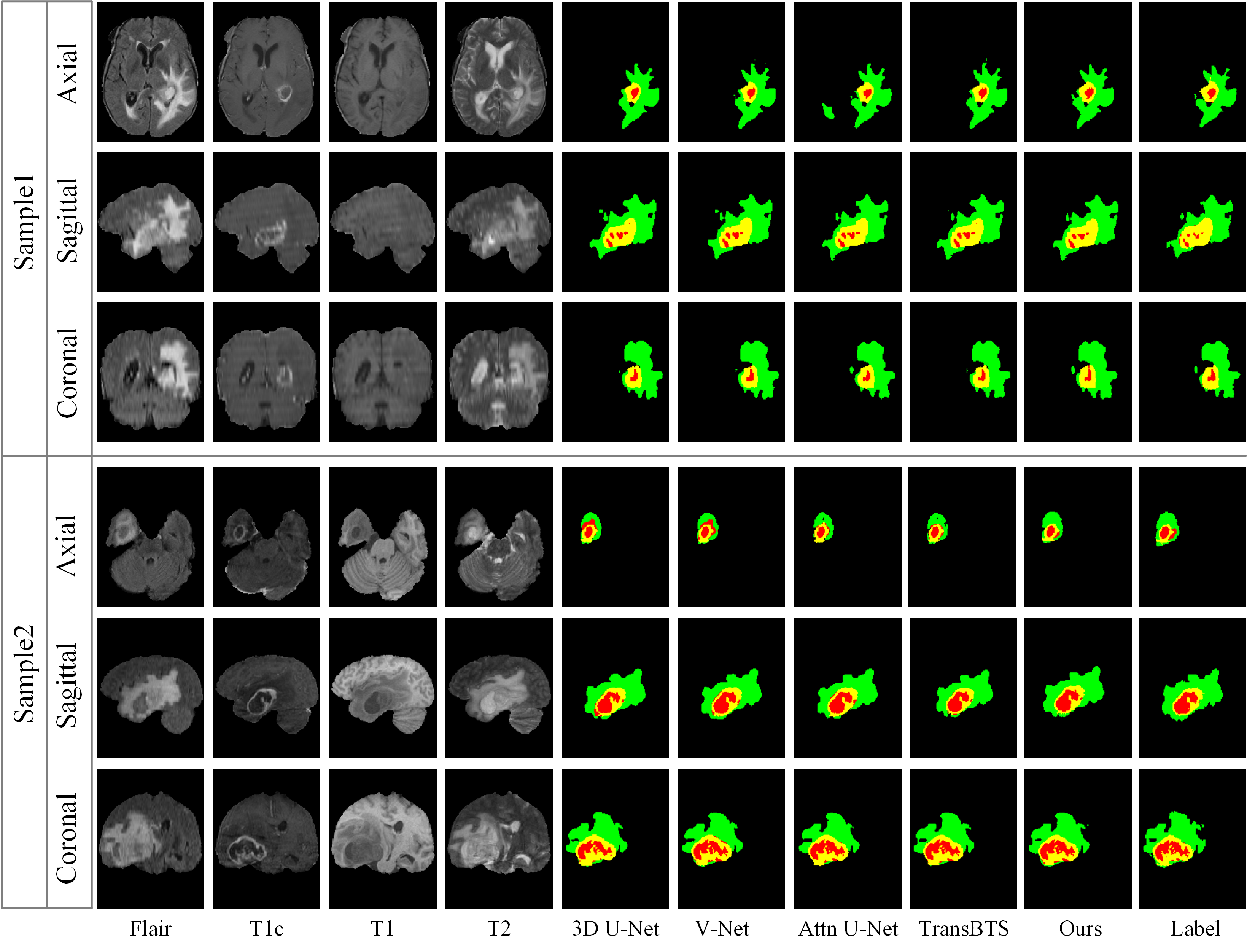}
	\caption{Visual comparison of the proposed method with different segmentation methods. The four left columns of images are the four input modalities, while the right shows the segmentation results and ground truths of different models. Segmentation results are shown from three perspectives: axial, sagittal and coronal for each sample.}
	\label{Fig7}
\end{figure*}

In addition, we also compare the proposed method with other state-of-the-art methods on BraTS2020. Similar to the above experiments, the same network hyper-parameter settings as BraTS2018 are used on this dataset. Experimental results of different methods on the BraTS2020 are shown in Table \ref{table3}. It can be seen that Dice coefficients of the proposed method in ET and WT regions are optimal, with $79.04\%$ and $90.05\%$, respectively. In TC and ET regions, HD95 distance also reaches optimal, i.e., 7.08mm and 26.57mm, respectively. Moreover, the average values of Dice coefficients and HD5 distance over the TC, ET and WT regions for the proposed method are optimal compared to those of the comparison methods.

In order to demonstrate the effectiveness of the proposed method, we visually compare the segmentation results. Limited by space, we only take the example of the segmentation results on BraTS2018. Since some state-of-the-arts have not published the code, only the results of 3D U-Net~\cite{Ahmed20163D}, V-Net~\cite{Milletari2016VNet}, Attn U-Net~\cite{Oktay2018Attention}, TransBTS~\cite{Wang2021TransBTS} and the method in this paper are visually compared. It is not possible to evaluate the segmentation quality by comparing the results in the validation set with the ground truth for the BraTS2018 validation set in which the ground truth is not available. To this end, we randomly re-divide the training set of BraTS2018 into the training set and test set by the ratio of 8:2. The segmentation results of the proposed method and 3D U-Net, V-Net, Attn U-Net, TransBTS are shown in Fig.~\ref{Fig7}. It can be seen that the proposed method can identify and locate the tumor area and boundary more accurately, while other results are more coarse.

\subsection{Ablation Study}
The proposed method is mainly composed of three core parts: CTP, KIIMI and PFRF. Since CTP is mainly composed of PFG and MII, we also reveal the effectiveness of MII. The key component of PFRF is PDM, whose effectiveness is verified in addition to KIIMI. In this process, we remove CTP, KIIMI and PFRF from the proposed model for the Baseline. The experimental results on BraTS2018 validation set are taken as an example for verification. Moreover, the 285 samples in the BraTS2018 training set are randomly re-divided into training samples and test samples according to 8:2 for comparison of different components. The results of the ablation experiment are shown in Table \ref{table4}. The results indicate that the incremental inclusion of proposed modules enables our full model architecture to achieve a highest average Dice score of 85.41\%. It also reduces the Hausdorff distance to $4.57$ mm.
 
\begin{figure*}[t]
	\centering
	\includegraphics[width=0.9\textwidth]{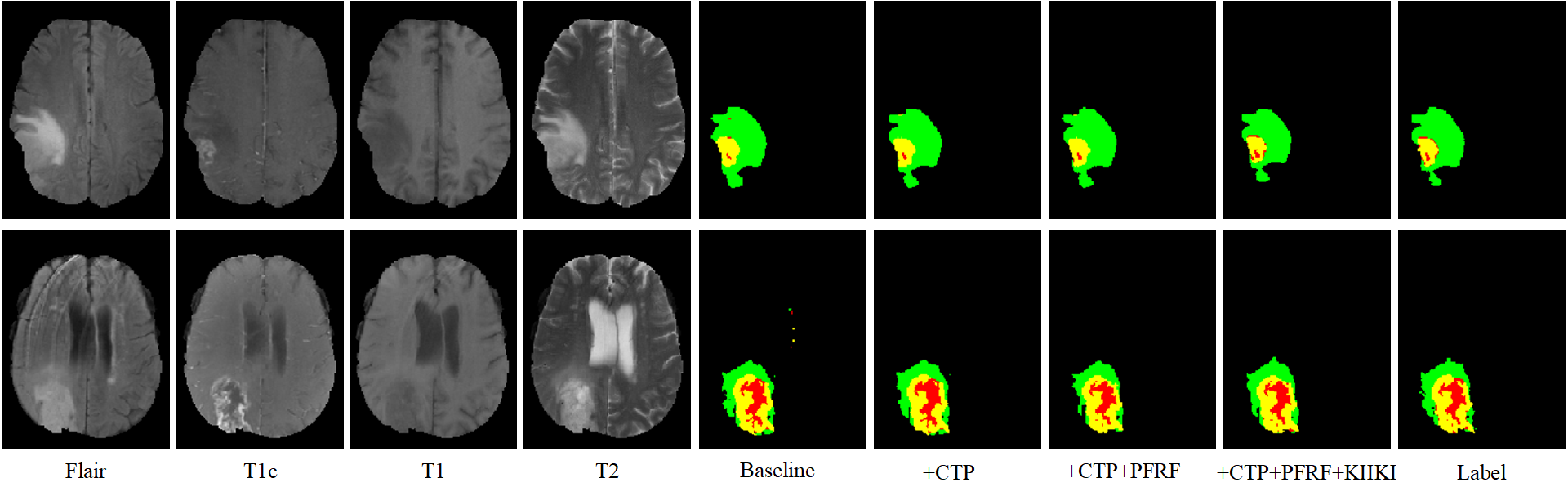}
	\caption{Visual comparison of the effectiveness of different components.}
	\label{Fig8}
\end{figure*}

\begin{figure}[t]
	\centering
	\includegraphics[width=0.9\linewidth]{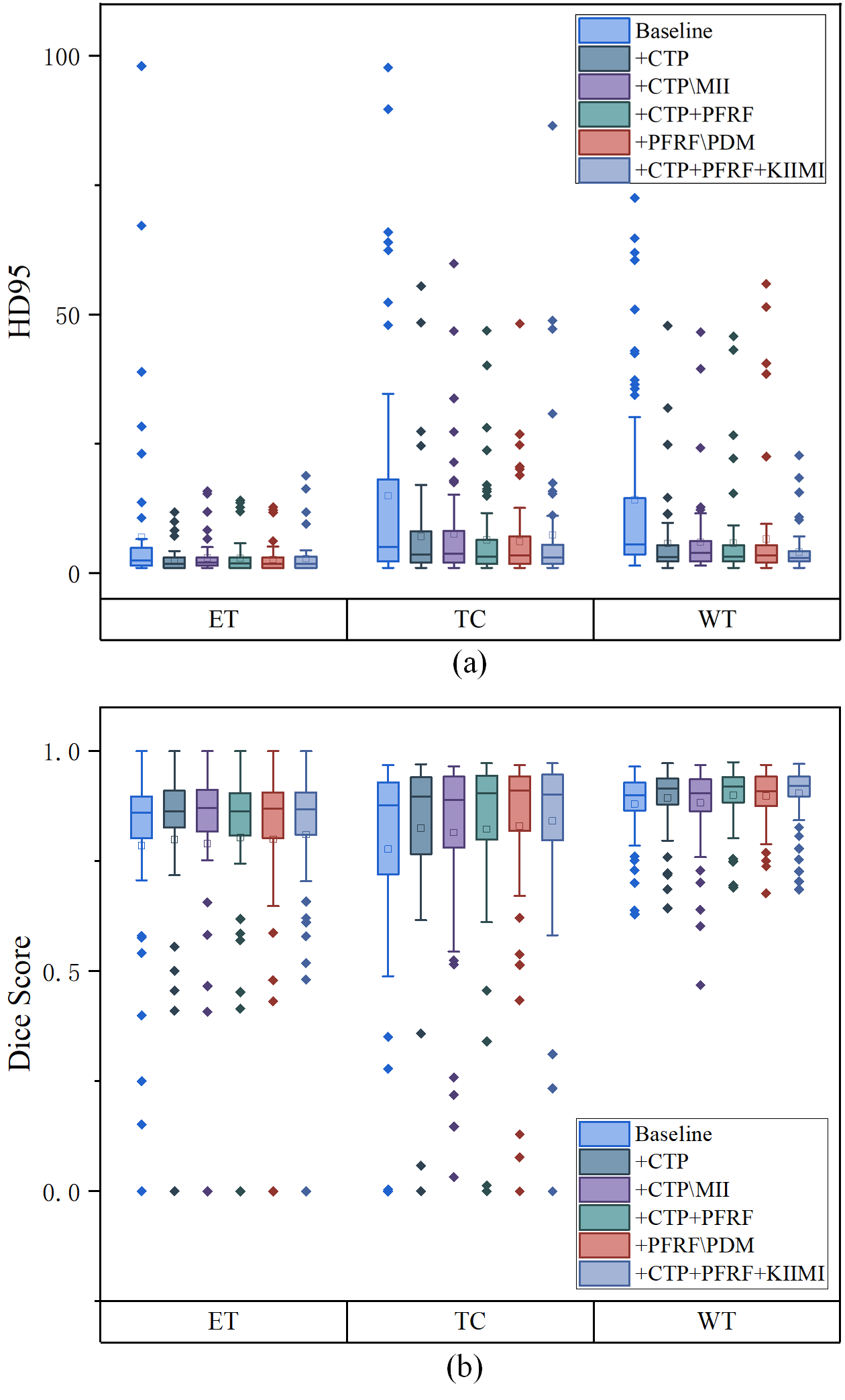}
	\caption{Box plot of the Dice score and Hausdorff Distance for different models in the ablation study. (a) Dice score; (b) Hausdorff Distance}
	\label{Fig9}
\end{figure}

Moreover,we visualize the segmentation results of different components as shown in Fig.~\ref{Fig8}. It can be observed that the complete model, compared to other combinations, achieves more refined segmentation, particularly in the border areas of the tumor. In addition, we have plotted box plots of the Dice Score and Hausdorff Distance for different models in the ablation study, as shown in Fig.~\ref{Fig9}. By examining these two box plots, we can draw the following conclusions. As we progressively added the proposed components, the models exhibited higher accuracy and smaller standard deviation in terms of Dice Score and Hausdorff Distance in the ET (Enhancing Tumor), WT (Whole Tumor), and TC (Tumor Core) regions compared to the baseline model.

\begin{table*}[htbp]\footnotesize
	\centering {\caption{Ablation study on BraTS2018 validation set. CTP\textbackslash MII indicates the removal of MII from CTP. PFRF\textbackslash PDM represents removal of PDM from PFRF. ``Avg." denotes the average value in the TC, ET and WT.}\label{table4}
		\renewcommand\arraystretch{1.0}
		{\footnotesize\centerline{\tabcolsep=12.0pt
				\begin{tabular*}{0.9\textwidth}{cc cccc cccc}
					\hline
					\multirow{2}*{Methods} & \multicolumn{4}{c}{Dice Score($\%$)$\uparrow$} & \multicolumn{4}{c}{HD95 (mm)$\downarrow$} \\
					\cline{2-5} \cline{7-10}
					&TC &ET &WT  &Avg. &&TC &ET &WT  &Avg.\\
					\hline
					Baseline &79.27 &78.58 &87.53 &81.79 &&8.40 &4.03 &7.41 &6.61\\
					+CTP &82.51 &79.87 &89.40 &83.93 &&7.06 &2.50 &5.71 &5.09\\
					+CTP\textbackslash MII &81.49 &78.97 &88.25 &82.90 &&7.48 &2.90 &5.96 &5.45\\
					+CTP+PFRF &82.51 &80.37 &90.11 &84.33 &&6.39 &2.87 &5.55 &4.94\\
					+CTP+PFRF\textbackslash PDM &83.00 &79.98 &89.79 &84.26 &&6.05 &2.61 &6.58  &5.08\\
					+CTP+PFRF+KIIMI &84.60 &81.06 &90.57 &85.41 &&6.88 &2.85 &3.97 &4.57\\
					\hline
	\end{tabular*}}}}
\end{table*}

\subsection{Discussion on effectiveness of CTP} To ascertain the efficiency of the Construction of Tumor Prototype (CTP), we juxtaposed the performances of ``Baseline+CTP" and the ``Baseline" model. As demonstrated in Table \ref{table4}, the Dice scores for "Baseline" on TC, ET, and WT reached $79.27\%$, $78.58\%$, and $87.53\%$ respectively, while the HD95 were $8.40$ mm, $4.03$~mm, and $7.41$~mm. Moreover, ``Baseline+CTP" achieved scores of $82.51\%$, $79.87\%$, and $89.40\%$ on TC, ET, and WT, while the HD95 improved to $7.06$~mm, $2.50$~mm, and $5.71$~mm. These results suggest a marked improvement in segmentation performance with significant gains in both Dice scores and Hausdorff distance metrics, validating the model's effective learning of the prototype feature representation of the tumor via the CTP process. This conclusion is further supported by the visual segmentation results presented in Fig.~\ref{Fig8}, which clearly demonstrate a noticeable reduction in segmentation error in certain tumor areas upon the addition of the CTP module. These observations collectively affirm the efficacy of CTP in enhancing tumor sub-region feature discrimination and subsequently improving segmentation performance. 

Additionally, the ``Baseline+CTP\textbackslash MII'' results emphasize the constructive role that multi-modal information interaction (MII) plays within the CTP framework in boosting segmentation performance. This experiment underscores the model's ability to build a more precise tumor prototype through the interchange of information among different modalities, thereby effectively facilitating the fusion of multi-modal features in subsequent tasks. The introduction of CTP has not only improved statistical metrics but has led to more practical and granular understandings of tumor structures.

\subsection{Discussion on effectiveness of PFRF} In our study, we propose a Prototype-Driven Feature Representation and Fusion (PFRF) technique to amplify features consistent with prototype categories. We examine data from Table \ref{table4} to qualitatively assess PFRF's effectiveness. The introduction of PFRF boosts Dice scores for ``Baseline+CTP+PFRF" from ($82.51\%$, $79.87\%$, $89.40\%$) to ($82.51\%$, $80.37\%$, $90.11\%$), compared to ``Baseline+CPT" on TC, ET, and WT. Concurrently, the HD95 values decrease from ($7.06$~mm, $2.50$~mm, $5.71$~mm) to ($6.39$~mm, $2.87$~mm, $5.55$~mm). These results signify that incorporating PFRF into our network significantly improves Dice scores and Hausdorff distance metrics. This supports our assertion that by embedding prototype features, the model effectively assimilates discriminative tumor characteristics. Enhancing features consistent with tumors strategically, we have significantly improved the model's segmentation precision. 

Furthermore, we analyzed the effectiveness of the Prototype-Driven Module (PDM) within PFRF. Upon its removal from PFRF, ``Baseline+CPT+PFRF\textbackslash PDM" undergoes an overall performance decline, highlighting the essential role of PDM in the PFRF framework. On closer inspection, we discovered that single-modal features, following the PDM operation, possess more discriminative tumor features than a direct convolution-based fusion. Therefore, they provide a richer information set advantageous for segmentation. Given this experimental context, it's vital to underscore the significant contribution of our proposed PFRF strategy to the model's overall performance by selectively enhancing relevant tumor features. The empirically determined improvements authenticate the importance of prototype features for efficient segmentation, demonstrating their unique role in enhancing the model's comprehension of tumor-specific characteristics. Additionally, the crucial role of the PDM within PFRF becomes evident in the performance deterioration upon its removal. The PDM's ability to enrich single-modal features with more discriminative tumor characteristics adds another layer of information, further refining the segmentation process. Collectively, these findings affirm our claim that the proposed PFRF strategy, supported by the PDM operation, significantly improves tumor segmentation tasks.

\subsection{Discussion on effectiveness of KIIMI} To further augment our model's segmentation capabilities, we developed a Key Information Integration Method from Multiple Experts (KIIMI). As depicted in Fig.~\ref{Fig8}, the substantial enhancement in the model's segmentation performance following the inclusion of this module, especially in non-enhanced tumor regions, attests to its effectiveness. As showcased in Table \ref{table4}, the incorporation of KIIMI propels an increase in the Dice scores of ``Baseline+CTP+PFRF+KIIMI" in the TC, ET, and WT regions by $2.09\%$, $0.69\%$, and $0.46\%$, respectively, in comparison to ``Baseline+CTP+PFRF". Simultaneously, HD95 values for ET and WT see reductions of $0.02$~mm and $1.58$~mm, respectively. These findings underline that KIIMI can notably enhance the classification and localization ability of the proposed method. The results further suggest that the integration of tumor features into the decoder significantly bolsters the model's segmentation performance, thereby generating more accurate and consistent predictions.

\section{Conclusion}
\label{sec:conclusion}
In this paper, we propose a tumor prototype-driven multi-modal brain tumor image segmentation method, which improves the discriminating ability of features in tumor regional through CTP, PFRF, and KIIMI, so as to obtain better segmentation results. In CTP, we not only consider the different sensitivities of different modalities to different regions of tumors, but also the diversity of tumor regions for the same category. A MII mechanism is designed to realize the mutual transmission of image information between different modalities, avoiding the defect of limited single-modal information. Prototype-driven feature representation uses tumor sub-region prototype to highlight the role of tumor specific sub-region features. KIIMI utilizes the features on different depth layers of the network to assist the segmentation of tumor, further improving the segmentation performance of the network. A large number of experimental results prove the effectiveness of the above method and its superiority over the existing methods.

\section*{References}
\bibliography{main}

\begin{thebibliography}{10}
\providecommand{\url}[1]{#1}
\csname url@samestyle\endcsname
\providecommand{\newblock}{\relax}
\providecommand{\bibinfo}[2]{#2}
\providecommand{\BIBentrySTDinterwordspacing}{\spaceskip=0pt\relax}
\providecommand{\BIBentryALTinterwordstretchfactor}{4}
\providecommand{\BIBentryALTinterwordspacing}{\spaceskip=\fontdimen2\font plus
\BIBentryALTinterwordstretchfactor\fontdimen3\font minus
  \fontdimen4\font\relax}
\providecommand{\BIBforeignlanguage}[2]{{%
\expandafter\ifx\csname l@#1\endcsname\relax
\typeout{** WARNING: IEEEtran.bst: No hyphenation pattern has been}%
\typeout{** loaded for the language `#1'. Using the pattern for}%
\typeout{** the default language instead.}%
\else
\language=\csname l@#1\endcsname
\fi
#2}}
\providecommand{\BIBdecl}{\relax}
\BIBdecl

\bibitem{Ding2017RFNet}
Y.~Ding, X.~Yu, and Y.~Yang, ``Rfnet: Region-aware fusion network for
  incomplete multi-modal brain tumor segmentation,'' in \emph{2021 IEEE/CVF
  International Conference on Computer Vision (ICCV)}, 2021, pp. 3955--3964.

\bibitem{Bauer2011Fully}
S.~Bauer, L.-P. Nolte, and M.~Reyes, ``Fully automatic segmentation of brain
  tumor images using support vector machine classification in combination with
  hierarchical conditional random field regularization,'' in
  \emph{International Conference on Medical Image Computing and
  Computer-Assisted Intervention--MICCAI 2011}, 2011, pp. 354--361.

\bibitem{9585643}
Y.~Ding, W.~Zheng, J.~Geng, Z.~Qin, K.-K.~R. Choo, Z.~Qin, and X.~Hou,
  ``Mvfusfra: A multi-view dynamic fusion framework for multimodal brain tumor
  segmentation,'' \emph{IEEE Journal of Biomedical and Health Informatics},
  vol.~26, no.~4, pp. 1570--1581, 2022.

\bibitem{Chen2019S3D}
W.~Chen, B.~Liu, S.~Peng, J.~Sun, and X.~Qiao, ``S3d-unet: Separable 3d u-net
  for brain tumor segmentation,'' in \emph{International MICCAI Brainlesion
  Workshop 2018}, 2019, pp. 358--368.

\bibitem{9793692}
H.~Fu, G.~Wang, W.~Lei, W.~Xu, Q.~Zhao, S.~Zhang, K.~Li, and S.~Zhang,
  ``Hmrnet: High and multi-resolution network with bidirectional feature
  calibration for brain structure segmentation in radiotherapy,'' \emph{IEEE
  Journal of Biomedical and Health Informatics}, vol.~26, no.~9, pp.
  4519--4529, 2022.

\bibitem{Zhuang2023A}
Y.~Zhuang, H.~Liu, E.~Song, and C.-C. Hung, ``A 3d cross-modality feature
  interaction network with volumetric feature alignment for brain tumor and
  tissue segmentation,'' \emph{IEEE Journal of Biomedical and Health
  Informatics}, vol.~27, no.~1, pp. 75--86, 2023.

\bibitem{Zikic2012Decision}
D.~Zikic, B.~Glocker, E.~Konukoglu, A.~Criminisi, C.~Demiralp, J.~Shotton,
  O.~M. Thomas, T.~Das, R.~Jena, and S.~J. Price, ``Decision forests for
  tissue-specific segmentation of high-grade gliomas in multi-channel mr,'' in
  \emph{International Conference on Medical Image Computing and
  Computer-Assisted Intervention--MICCAI 2012}, 2012, pp. 369--376.

\bibitem{Saueressig2020Exploring}
C.~Saueressig, A.~Berkley, E.~Kang, R.~Munbodh, and R.~Singh, ``Exploring
  graph-based neural networks for automatic brain tumor segmentation,'' in
  \emph{DataMod 2020: From Data to Models and Back}, 2020, pp. 18--37.

\bibitem{Sun2020A}
L.~Sun, W.~Ma, X.~Ding, Y.~Huang, D.~Liang, and J.~Paisley, ``A 3d spatially
  weighted network for segmentation of brain tissue from mri,'' \emph{IEEE
  Transactions on Medical Imaging}, vol.~39, no.~4, pp. 898--909, 2020.

\bibitem{Myronenko20193D}
A.~Myronenko, ``3d mri brain tumor segmentation using autoencoder
  regularization,'' in \emph{International MICCAI Brainlesion Workshop 2018},
  2019, pp. 311--320.

\bibitem{Kong2018Hybrid}
X.~Kong, G.~Sun, Q.~Wu, J.~Liu, and F.~Lin, ``Hybrid pyramid u-net model for
  brain tumor segmentation,'' in \emph{IIP 2018: Intelligent Information
  Processing IX}, 2018, pp. 346--355.

\bibitem{Wang2021TransBTS}
W.~Wang, C.~Chen, M.~Ding, H.~Yu, S.~Zha, and J.~Li, ``Transbts: Multimodal
  brain tumor segmentation using transformer,'' in \emph{International
  Conference on Medical Image Computing and Computer Assisted
  Intervention--MICCAI 2021}, 2021, pp. 109--119.

\bibitem{Hatamizadeh2022Swin}
A.~Hatamizadeh, V.~Nath, and e.~Yucheng~Tang, ``Swin unetr: Swin transformers
  for semantic segmentation of brain tumors in mri images,'' in
  \emph{International MICCAI Brainlesion Workshop 2021}, 2022, pp. 272--284.

\bibitem{Hatamizadeh2022UNETR}
A.~Hatamizadeh, Y.~Tang, and e.~Nath, Vishwesh, ``Unetr: Transformers for 3d
  medical image segmentation,'' in \emph{2022 IEEE/CVF Winter Conference on
  Applications of Computer Vision (WACV)}, 2022, pp. 1748--1758.

\bibitem{Zhou2022A}
T.~Zhou, S.~Ruan, P.~Vera, and S.~Canu, ``A tri-attention fusion guided
  multi-modal segmentation network,'' \emph{Pattern Recognition}, vol. 124, p.
  108417, 2022.

\bibitem{Zhang2020Exploring}
D.~Zhang, G.~Huang, Q.~Zhang, J.~Han, J.~Han, Y.~Wang, and Y.~Yu, ``Exploring
  task structure for brain tumor segmentation from multi-modality mr images,''
  \emph{IEEE Transactions on Image Processing}, vol.~29, pp. 9032--9043, 2020.

\bibitem{Havaei2017Brain}
M.~Havaei, A.~Davy, D.~Warde-Farley, A.~Biard, A.~Courville, Y.~Bengio, C.~Pal,
  P.-M. Jodoin, and H.~Larochelle, ``Brain tumor segmentation with deep neural
  networks,'' \emph{Medical Image Analysis}, vol.~35, pp. 18--31, 2017.

\bibitem{Chen2018Focus}
X.~Chen, J.~H. Liew, W.~Xiong, C.-K. Chui, and S.-H. Ong, ``Focus, segment and
  erase: An efficient network for multi-label brain tumor segmentation,'' in
  \emph{European Conference on Computer Vision}, 2018, pp. 674--689.

\bibitem{Wang2019Automatic}
G.~Wang, W.~Li, S.~Ourselin, and T.~Vercauteren, ``Automatic brain tumor
  segmentation using convolutional neural networks with test-time
  augmentation,'' in \emph{International MICCAI Brainlesion Workshop 2018},
  2019, pp. 61--72.

\bibitem{Rezaei2019voxel}
M.~Rezaei, H.~Yang, and C.~Meinel, ``voxel-gan: Adversarial framework for
  learning imbalanced brain tumor segmentation,'' in \emph{International MICCAI
  Brainlesion Workshop 2018}, 2019, pp. 321--333.

\bibitem{Xing2022NestedFormer}
Z.~Xing, L.~Yu, L.~Wan, T.~Han, and L.~Zhu, ``Nestedformer: Nested
  modality-aware transformer for brain tumor segmentation,'' in \emph{Medical
  Image Computing and Computer Assisted Intervention-MICCAI 2022}, 2022, pp.
  140--150.

\bibitem{Kamnitsas2017Efficient}
K.~Kamnitsas, C.~Ledig, V.~F. Newcombe, J.~P. Simpson, A.~D. Kane, D.~K. Menon,
  D.~Rueckert, and B.~Glocker, ``Efficient multi-scale 3d cnn with fully
  connected crf for accurate brain lesion segmentation,'' \emph{Medical Image
  Analysis}, vol.~36, pp. 61--78, 2017.

\bibitem{Vaswani2017Attention}
A.~Vaswani, N.~Shazeer, N.~Parmar, J.~Uszkoreit, L.~Jones, A.~N. Gomez,
  L.~Kaiser, and I.~Polosukhin, ``Attention is all you need,'' in
  \emph{NeurIPS}, 2017, pp. 5998--6008.

\bibitem{Chen2019OctopusNet}
Y.~Chen, J.~Chen, D.~Wei, Y.~Li, and Y.~Zheng, ``Octopusnet: A deep learning
  segmentation network for multi-modal medical images,'' in \emph{MMMI 2019:
  Multiscale Multimodal Medical Imaging}, 2019, pp. 17--25.

\bibitem{Zhang2021Cross}
D.~Zhang, G.~Huang, Q.~Zhang, J.~Han, J.~Han, and Y.~Yu, ``Cross-modality deep
  feature learning for brain tumor segmentation,'' \emph{Pattern Recognition},
  vol. 110, p. 107562, 2021.

\bibitem{Zhou2021Latent}
T.~Zhou, S.~Canu, P.~Vera, and S.~Ruan, ``Latent correlation representation
  learning for brain tumor segmentation with missing mri modalities,''
  \emph{IEEE Transactions on Image Processing}, vol.~30, pp. 4263--4274, 2021.

\bibitem{Zhang2022mmFormer}
Y.~Zhang, N.~He, J.~Yang, Y.~Li, D.~Wei, Y.~Huang, Y.~Zhang, Z.~He, and
  Y.~Zheng, ``mmformer: Multimodal medical transformer for incomplete
  multimodal learning of brain tumor segmentation,'' in \emph{International
  Conference on Medical Image Computing and Computer Assisted
  Intervention--MICCAI 2022}, 2022, pp. 107--117.

\bibitem{Zhang2021Modality}
Y.~Zhang, J.~Yang, J.~Tian, Z.~Shi, C.~Zhong, Y.~Zhang, and Z.~He,
  ``Modality-aware mutual learning for multi-modal medical image
  segmentation,'' in \emph{International Conference on Medical Image Computing
  and Computer Assisted Intervention-MICCAI 2021}, 2021, pp. 589--599.

\bibitem{Chen2019Robust}
C.~Chen, Q.~Dou, Y.~Jin, H.~Chen, J.~Qin, and P.-A. Heng, ``Robust multimodal
  brain tumor segmentation via feature disentanglement and gated fusion,'' in
  \emph{International Conference on Medical Image Computing and Computer
  Assisted Intervention--MICCAI 2019}, 2019, pp. 447--456.

\bibitem{Menze2015Multimodal}
B.~H. Menze, A.~Jakab, S.~Bauer, and et~al., ``The multimodal brain tumor image
  segmentation benchmark (brats),'' \emph{IEEE Transactions on Medical
  Imaging}, vol.~34, no.~10, pp. 1993--2024, 2015.

\bibitem{Dice1945Measures}
L.~R. Dice, ``Measures of the amount of ecologic association between species,''
  \emph{Ecology}, vol.~26, no.~3, pp. 297--302, 1945.

\bibitem{Dorent2019Hetero}
R.~Dorent, S.~Joutard, M.~Modat, S.~Ourselin, and T.~Vercauteren,
  ``Hetero-modal variational encoder-decoder for joint modality completion and
  segmentation,'' in \emph{International Conference on Medical Image Computing
  and Computer Assisted Intervention-- MICCAI 2019}, 2019, pp. 74--82.

\bibitem{Paszke2019PyTorch}
A.~Paszke, S.~Gross, F.~Massa, A.~Lerer, and et~al., ``Pytorch: an imperative
  style, high-performance deep learning library,'' in \emph{NIPS'19}, 2019, pp.
  8026--8037.

\bibitem{Kingma2014Adam}
D.~P. Kingma and J.~Ba, ``Adam: A method for stochastic optimization,'' in
  \emph{International Conference on Learning Representations}, 2014, pp.
  109--119.

\bibitem{Milletari2016VNet}
F.~Milletari, N.~Navab, and S.-A. Ahmadi, ``V-net: Fully convolutional neural
  networks for volumetric medical image segmentation,'' in \emph{2016 Fourth
  International Conference on 3D Vision (3DV)}, 2016, pp. 565--571.

\bibitem{Ahmed20163D}
\"{O}zg\"{u}n {\c{C}}i{\c{c}}ek, A.~Abdulkadir, S.~S. Lienkamp, T.~Brox, and
  O.~Ronneberger, ``3d u-net: Learning dense volumetric segmentation from
  sparse annotation,'' in \emph{International Conference on Medical Image
  Computing and Computer-Assisted Intervention--MICCAI 2016}, 2016, pp.
  424--432.

\bibitem{Salehi2017Auto}
S.~S.~M. Salehi, D.~Erdogmus, and A.~Gholipour, ``Auto-context convolutional
  neural network (auto-net) for brain extraction in magnetic resonance
  imaging,'' \emph{IEEE transactions on medical imaging}, vol.~36, no.~11, pp.
  2319--2330, 2017.

\bibitem{Weninger2019Segmentation}
L.~Weninger, O.~Rippel, S.~Koppers, and D.~Merhof, ``Segmentation of brain
  tumors and patient survival prediction: Methods for the brats 2018
  challenge,'' in \emph{International MICCAI Brainlesion Workshop 2018}, 2019,
  pp. 3--12.

\bibitem{Liu2021CANet}
Z.~Liu, L.~Tong, L.~Chen, F.~Zhou, Z.~Jiang, Q.~Zhang, Y.~Wang, C.~Shan, L.~Li,
  and H.~Zhou, ``Canet: Context aware network for brain glioma segmentation,''
  \emph{IEEE Transactions on Medical Imaging}, vol.~40, no.~7, pp. 1763--1777,
  2021.

\bibitem{Wang2019AST}
P.~Wang, S.~Liu, and J.~Peng, ``Ast-net: Lightweight hybrid transformer for
  multimodal brain tumor segmentation,'' in \emph{2022 26th International
  Conference on Pattern Recognition (ICPR)}, 2022, pp. 4623--4629.

\bibitem{Chang2023DPAFNet}
Y.~Chang, Z.~Zheng, Y.~Sun, M.~Zhao, Y.~Lu, and Y.~Zhang, ``Dpafnet: A residual
  dual-path attention-fusion convolutional neural network for multimodal brain
  tumor segmentation,'' \emph{Biomedical Signal Processing and Control},
  vol.~79, p. 104037, 2023.

\bibitem{Oktay2018Attention}
O.~Oktay, J.~Schlemper, L.~L. Folgoc, M.~Lee, M.~Heinrich, K.~Misawa, K.~Mori,
  S.~McDonagh, N.~Y. Hammerla, B.~Kainz, B.~Glocker, and D.~Rueckert,
  ``Attention u-net: Learning where to look for the pancreas,'' \emph{arXiv
  preprint}, 2018.

\bibitem{Li2020Multi}
X.~Li, G.~Luo, and K.~Wang, ``Multi-step cascaded networks for brain tumor
  segmentation,'' in \emph{International MICCAI Brainlesion Workshop 2019},
  2020, pp. 163--173.

\bibitem{3du-net19}
F.~Wang, R.~Jiang, L.~Zheng, C.~Meng, and B.~B. Biswal, ``3d u-net based brain
  tumor segmentation and survival days prediction,'' in \emph{International
  MICCAI Brainlesion Workshop}, 2019, pp. 131--141.

\bibitem{Cao2021Swin}
H.~Cao, Y.~Wang, J.~Chen, D.~Jiang, X.~Zhang, Q.~Tian, and M.~Wang,
  ``Swin-unet: Unet-like pure transformer for medical image segmentation,''
  \emph{arXiv preprint}, 2021.

\bibitem{Li2022TransBTSV2}
J.~Li, W.~Wang, C.~Chen, T.~Zhang, S.~Zha, J.~Wang, and H.~Yu, ``Transbtsv2:
  Towards better and more efficient volumetric segmentation of medical
  images,'' \emph{arXiv preprint}, 2022.

\bibitem{Chen2021TransUNet}
J.~Chen, Y.~Lu, Q.~Yu, X.~Luo, E.~Adeli, Y.~Wang, L.~Lu, A.~L. Yuille, and
  Y.~Zhou, ``Transunet: Transformers make strong encoders for medical image
  segmentation,'' \emph{arXiv preprint}, 2021.

\bibitem{Valanarasu2022KiU}
J.~M.~J. Valanarasu, V.~A. Sindagi, I.~Hacihaliloglu, and V.~M. Patel,
  ``Kiu-net: Overcomplete convolutional architectures for biomedical image and
  volumetric segmentation,'' \emph{IEEE Transactions on Medical Imaging},
  vol.~41, no.~4, pp. 965--976, 2022.

\bibitem{Li2022Category}
J.~Li, H.~Yu, C.~Chen, M.~Ding, and S.~Zha, ``Category guided attention network
  for brain tumor segmentation in mri,'' \emph{Physics in Medicine \& Biology},
  vol.~67, no.~8, p. 085014, 2022.

\bibitem{Zhang2018Road}
Z.~Zhang, Q.~Liu, and Y.~Wang, ``Road extraction by deep residual u-net,''
  \emph{IEEE Geoscience and Remote Sensing Letters}, vol.~15, no.~5, pp.
  749--753, 2018.

\bibitem{Zhou2019ModGen}
Z.~Zhou, V.~Sodha, M.~M.~R. Siddiquee, R.~Feng, N.~Tajbakhsh, M.~B. Gotway, and
  J.~Liang, ``Models genesis: Generic autodidactic models for 3d medical image
  analysis,'' in \emph{International Conference on Medical Image Computing and
  Computer-Assisted Intervention (MICCAI)}, 2019, pp. 384--393.

\bibitem{Liu2023M3AE}
H.~Liu, D.~Wei, D.~Lu, J.~Sun, L.~Wang, and Y.~Zheng, ``M3ae: Multimodal
  representation learning for brain tumor segmentation with missing
  modalities,'' in \emph{The 37th AAAI Conference on Artificial Intelligence},
  2023.

\bibitem{Luo2021HDCNet}
Z.~Luo, Z.~Jia, Z.~Yuan, and J.~Peng, ``Hdc-net: Hierarchical decoupled
  convolution network for brain tumor segmentation,'' \emph{IEEE Journal of
  Biomedical and Health Informatics}, vol.~25, no.~3, pp. 737--745, 2021.

\bibitem{Ding2021RFNet}
Y.~Ding, X.~Yu, and Y.~Yang, ``Rfnet: Region-aware fusion network for
  incomplete multi-modal brain tumor segmentation,'' in \emph{IEEE/CVF
  International Conference on Computer Vision (ICCV)}, 2021, pp. 3955--3964.

\bibitem{Wang2021ACN}
Y.~Wang, Y.~Zhang, Y.~Liu, Z.~Lin, J.~Tian, C.~Zhong, Z.~Shi, J.~Fan, and
  Z.~He, ``Acn: Adversarial co-training network for brain tumor segmentation
  with missing modalities,'' in \emph{International Conference on Medical Image
  Computing and Computer-Assisted Intervention (MICCAI)}, 2021, pp. 410--420.

\bibitem{Azad2022SMUNet}
R.~Azad, N.~Khosravi, and D.~Merhof, ``Smu-net: Style matching u-net for brain
  tumor segmentation with missing modalities,'' \emph{arXiv preprint}, 2022.

\end{thebibliography}
\end{document}